\documentclass[%
 reprint,
 amsmath,amssymb,
 aps,
superscriptaddress,
]{revtex4-2}

\usepackage[dvipsnames]{xcolor}
\usepackage{graphicx}
\usepackage{dcolumn}
\usepackage{bm}
\usepackage{lipsum}

\usepackage{url}
\usepackage[breaklinks]{hyperref}

\usepackage[dvipsnames]{xcolor}
\definecolor{mblue}{RGB}{42, 54, 144} 
\hypersetup{colorlinks=true,breaklinks=true,urlcolor=blue,linkcolor=blue,citecolor=blue}

\usepackage{placeins}
\usepackage[mathlines]{lineno}
\usepackage{siunitx}
\usepackage{braket}

\DeclareSIUnit\gauss{G}
\usepackage{tikz}

\definecolor{verdealberto}{rgb}{0.05, 0.64, 0.09}

\definecolor{chunkit_color}{rgb}{0.85, 0.75, 0.46}

\begin{document}


\title{Dipole-Mode Spectrum and Hydrodynamic Crossover in a Resonantly Interacting Two-Species Fermion Mixture}

\author{Zhu-Xiong Ye}
\affiliation{Institut f{\"u}r Experimentalphysik, Universit{\"a}t Innsbruck, Austria} 
\affiliation{State Key Laboratory of Quantum Optics Technologies and Devices, Institute of Opto-Electronics, Shanxi University, Taiyuan, 030006, China}

\author{Alberto Canali}
\affiliation{Institut f{\"u}r Experimentalphysik, Universit{\"a}t Innsbruck, Austria} 

\author{Chun-Kit Wong}
\affiliation{Institut f{\"u}r Experimentalphysik, Universit{\"a}t Innsbruck, Austria}
\affiliation{Institut f{\"u}r Quantenoptik und Quanteninformation (IQOQI), {\"O}sterreichische Akademie der Wissenschaften, Innsbruck, Austria}

\author{Marian Kreyer}
\affiliation{Institut f{\"u}r Experimentalphysik, Universit{\"a}t Innsbruck, Austria} 
\affiliation{Institut f{\"u}r Quantenoptik und Quanteninformation (IQOQI), {\"O}sterreichische Akademie der Wissenschaften, Innsbruck, Austria}

\author{Emil Kirilov}
\affiliation{Institut f{\"u}r Experimentalphysik, Universit{\"a}t Innsbruck, Austria} 
\affiliation{Institut f{\"u}r Quantenoptik und Quanteninformation (IQOQI), {\"O}sterreichische Akademie der Wissenschaften, Innsbruck, Austria}

\author{Rudolf Grimm}
\affiliation{Institut f{\"u}r Experimentalphysik, Universit{\"a}t Innsbruck, Austria} 
\affiliation{Institut f{\"u}r Quantenoptik und Quanteninformation (IQOQI), {\"O}sterreichische Akademie der Wissenschaften, Innsbruck, Austria}

\date{\today}

\begin{abstract}
Ultracold quantum-gas mixtures of fermionic atoms with resonant control of  interactions offer a unique test-bed to explore few- and many-body quantum states with unconventional properties. 
The emergence of such strongly correlated systems, as for instance symmetry-broken superfluids, is usually accompanied by hydrodynamic collective behavior. Thus, experimental progress in this field naturally requires a deep understanding of hydrodynamic regimes. Here, we report on experiments employing a tunable Fermi-Fermi mixture of $^{161}$Dy and $^{40}$K near quantum degeneracy. We investigate the full spectrum of dipole modes across a Feshbach resonance and characterize the crossover from collisionless to deep hydrodynamic behavior in measurements of frequencies and damping rates. We compare our results with a theoretical model that considers the motion of the mass centers of the two species and we identify the contributions of friction and mean-field interaction.
We show that one oscillating mode exists over the whole range of interactions, exhibiting striking changes of frequency and damping in the deep hydrodynamic regime. We observe the second oscillating mode to split into two purely exponential damping modes. One of these exponential modes shows very fast damping, faster than any other relevant timescale, and is largely insensitive against experimental imperfections. It provides an accurate measure for the interspecies drag effect, which generalizes the concept of spin drag explored in other experiments. {We finally characterize the interspecies drag in terms of a microscopic friction coefficient} and we discuss its unitarity-limited universal behavior on top of the resonance.
\end{abstract}

\maketitle

\section{Introduction}\label{sec:intro}

More than two decades of research on ultracold fermionic quantum gases 
have yielded a wealth of exciting insights into quantum matter featuring strong interactions~\cite{Varenna2006book, Varenna2014book, Varenna2022book, Zwerger2012tbb, Strinati2018tbb}. Two-component 
mixtures are commonly used in the experiments to enable $s$-wave interaction, which facilitates efficient cooling and allows to implement strong tunable interactions near Feshbach resonances \cite{Chin2010fri}.
While spin mixtures of fermionic atoms of the same species are routinely employed in many present experiments, mixtures composed of different fermionic species are much less established, presumably owing to their complex interaction properties and additional challenges in the preparation process.
However, fermionic species mixtures hold great promise for future experiments, enabling the exploration of novel phenomena through the introduction of mass imbalance as an additional degree of freedom.
Systems of fermions with constituents of different masses are at the heart of new phenomena in the few-body sector \cite{Efimov1973elo, Kartavtsev2007let, Nishida2008ufg, Blume2010bou, Bazak2017fbe, Mathy2011tma, Levinsen2009ads, Jag2014ooa, Naidon2017epa, Greene2017ufb}, regarding the physics of quantum impurities \cite{Mathy2011tma, Baroni2024mib, Massignan2025pia}, and in relation to unconventional pairing mechanisms \cite{Liu2003igs, Forbes2005scf, Iskin2006tsf, Parish2007pfc, Baranov2008spb, Gezerlis2009hlf, Gubbels2009lpi, Baarsma2013isp, Gubbels2013ifg, Wang2017eeo, Pini2021bmf} and the emergence of novel superfluids \cite{Bennemann2014both, Liu2023qsi}. A particular example and motivation for our {work on two-species fermion mixtures} is the prospect to observe the elusive Fulde-Ferrell-Larkin-Ovchinnikov (FFLO) state~\cite{Fulde1964sia,Larkin1964nss,Radzihovsky2010ifr}, which in 
{the case of mass imbalance} can be expected \cite{Gubbels2009lpi, Baarsma2013isp, Gubbels2013ifg, Wang2017eeo, Pini2021bmf} under realistically attainable experimental conditions.

The main requirements for experiments towards novel few- or many-body quantum states in fermionic mixtures include efficient cooling into the quantum-degenerate regime, the existence of sufficiently broad Feshbach resonance for interaction tuning, and collisional stability of the mixture in the resonance regime.
To our best knowledge, five different two-species fermion mixtures have been realized so far:
$^{6}$Li-$^{40}$K~\cite{Taglieber2008qdt, Wille2008eau, Voigt2009uhf,
Trenkwalder2011heo, Naik2011fri},
$^{161}$Dy-$^{40}$K~\cite{Ravensbergen2018poa,Ravensbergen2020rif, Ye2022ool, Soave2023otf}, 
$^{6}$Li-$^{53}$Cr~\cite{Neri2020roa, Ciamei2022euc, Ciamei2022ddf, Finelli2024ula}, 
$^{6}$Li-$^{167}$Er~\cite{Schafer2023oof} and 
$^{6}$Li-$^{173}$Yb~\cite{Hara2011qdm, Green2020fri}.
Among these mixtures the Dy-K system features a unique combination of favorable properties. It is rather straightforward to cool the mixture deep into the doubly degenerate regime, following standard procedures of laser cooling and trapping and subsequent evaporative cooling \cite{Ravensbergen2018poa}. The mixture features an ample spectrum of interspecies Feshbach resonances in the lowest spin channel, which is immune against two-body losses. The observed resonances include a broad one centered at a magnetic field of 217\,G \cite{Ravensbergen2020rif} and a bunch of low-field Feshbach resonances \cite{Ye2022ool} below 10\,G. As a particularly interesting candidate for interaction tuning and molecule formation we have {identified} a resonance near 7.3\,G \cite{Ye2022ool, Soave2023otf}. {
This isolated resonance lies in a magnetic-field region not contaminated by ultra-narrow interspecies and intraspecies \cite{Soave2022lff} resonances and it facilitates fast interaction control by relatively small, but not too small changes of the magnetic fields.} For resonant atomic Dy-K mixtures, we have observed long lifetimes close to one second and measured low three-body rate coefficients, indicating a substantial suppression of recombination losses 
\footnote{Although the interaction properties in the Dy-K mixture sound almost ideal, it should be mentioned that the extremely dense spectrum of intraspecies $^{161}$Dy resonances \cite{Burdick2016lls, Soave2022lff} causes problems when the magnetic field is to be ramped from the preparation field to the resonance of interest for the experiments and thus requires carefully optimized
ramping protocols.}.

Strongly interacting Fermi gases exhibit hydrodynamic behavior \cite{Varenna2006book, Giorgini2008tou}. Related phenomena have been observed in many experiments on spin mixtures, mainly studying the expansion dynamics \cite{Giorgini2008tou, Ohara2002ooa}
and collective modes \cite{Kinast2004efs, Bartenstein2004ceo, Altmeyer2007doa, Wright2007ftc, Nascimbene2009coo, Tey2013cmi}.
The dependence of such phenomena on the interaction strength and temperature provides information on the properties of the system and elucidates the basic mechanisms leading to hydrodynamics.
In fermionic species mixtures, only few observations of hydrodynamic behavior have been reported so far. The anisotropic expansion of a $^6$Li-$^{40}$K mixture after release from the trap \cite{Trenkwalder2011heo} was demonstrated as an effect already well known from spin mixtures \cite{Ohara2002ooa}. 
An interesting class of phenomena, distinct from those seen in spin mixtures, emerges when species with markedly different properties are hydrodynamically coupled, leading to a composite fluid with novel characteristics.
Manifestations of such joint behavior have been observed in the expansion of the $^{161}$Dy-$^{40}$K mixture \cite{Ravensbergen2020rif} and in collective oscillations of the $^6$Li-$^{53}$Cr mixture \cite{Finelli2024ula}.

Here, we probe hydrodynamic behavior in an optically trapped Dy-K Fermi-Fermi mixture by characterizing the full spectrum of {two-species dipole modes} across a Feshbach resonance. Such modes (Sec.~\ref{sec:modespectrum}) can be understood as the response of the harmonically trapped mixture to a small displacement between the two components, and they can be conveniently excited by application of a species-dependent force (Sec.~\ref{sec:experimental}). 
In spin mixtures, the corresponding {spin dipole mode} has been observed in Refs.~\cite{Sommer2011ust, Sommer2011sti, Valtolina2017etf} and analyzed to characterize universal spin transport {and ferromagnetic correlations}. In species mixtures, the behavior is generally richer because the uncoupled oscillations of the two components naturally take place with different frequencies. This lifts a degeneracy that is normally present in optically trapped spin mixtures and there leads to a coexistence of the spin dipole mode with an undamped center-of-mass oscillation. In contrast, in interacting species mixtures, all modes are damped and exhibit complex-valued eigenfrequencies. Wide tuning of the interspecies $s$-wave interaction allows us to realize the full spectrum of dipole modes in the crossover from collisionless to hydrodynamic behavior. Below a critical friction strength, the spectrum consists of two damped oscillation modes. 
For stronger dissipative coupling, one of these modes splits into two pure exponential damping modes, whereas the other one (`crossover mode') survives with a substantially shifted eigenfrequency.

We show that all our observations (Sec.~\ref{sec:results}) can be described quantitatively within a simple model considering the center-of-mass motion of both species together with a dissipative coupling effect (friction) and a reactive coupling (mean-field interaction). Fitting the model to the experimental results, we extract the values for the corresponding coupling constants and analyze their resonance behavior. We find our experimental results on all modes in full agreement with the theoretical model, which gives a complete picture of the crossover from collisionless to hydrodynamic behavior over a wide interaction range.

Deep in the hydrodynamic regime, the two exponential modes exhibit vastly different damping rates. While the slow mode corresponds to a strongly overdamped motion in position space, the fast mode essentially reflects a drag effect that, analogously to spin drag in a spin mixture, damps the relative motion in momentum space on a very short time scale. We find that the latter mode offers great practical advantages for accurate measurements, since it appears on a very short time scale, where other systematic effects like heating or interaction-induced deformations of the atomic clouds are negligible. 

We finally go beyond the macroscopic description of the drag effect between the two species (Sec.~\ref{sec:microfriction}), introducing a microscopic friction coefficient {to describe the local interspecies friction effect separated from the global trap dynamics. We extract experimental values for this coefficient} from our measurements of the fast damping mode. In our analysis we pay particular attention to the unitarity-limited regime on top of the resonance.
{This shows how our results are linked to a universal description of friction in a resonant two-component Fermi gas near quantum degeneracy and facilitates a comparison with other systems.}

The detailed understanding of hydrodynamic behavior represents an essential prerequisite (Sec.~\ref{sec:conclusion}) for future experiments exploring fermionic systems in interaction regimes beyond the present state of the art, such as novel superfluids.

\section{Two-species dipole mode spectrum}\label{sec:modespectrum}

\subsection{Model}
\label{ssec:model}

The basic idea to probe hydrodynamic behavior in a two-component Fermi gas by a dipole mode goes back to early theoretical work~\cite{Vichi1999coo}, which considered collisional relaxation of the spin-dipole mode in a balanced spin mixture of fermions.
To describe the spectrum of dipole modes in a harmonically trapped two-species quantum gas, we adopt a model from the literature.  In Refs.~\cite{Gensemer2001tfc, Ferlaino2003doi} 
the model was applied to imbalanced situations (spin and species mixtures) with different populations and trap frequencies. In further theoretical work \cite{Chiacchiera2010doi, Asano2020doo}, a reactive term was added to the model to describe the attractive or repulsive mean-field interaction between the two components. The approach is intuitive and straightforward on a phenomenological basis. Formally, it can be derived from the Boltzmann equation as described in some detail in Ref.~\cite{Chiacchiera2010doi}, see also Refs.~\cite{Vichi1999coo, Narushima2018das, Asano2020doo}.
{Solutions can also be obtained by numerical simulations \cite{Lepers2010nso, Goulko2012bes}.}

The model considers the one-dimensional motion of the mass centers of the two species. {We assume rigid mass distributions,} neglecting all dynamical effects related to the size and shape of the clouds. It contains three contributions to the forces on the atomic clouds, (1) the species-dependent restoring force of the harmonic trap, (2) a friction term for the relative motion, and (3) a reactive term for the mean-field interaction between the two components. All corresponding terms are linear, based on the assumptions of small displacements (smaller than the cloud sizes) and small relative velocities (smaller than typical velocities within the clouds).

Within these assumptions, the equations of motion (see App.~\ref{app:eqmotion} for more details) can be written as
a system of second-order linear differential equations
for the center-of-mass positions $y_{\rm Dy}$ and $y_{\rm K}$  of the two clouds:
\begin{subequations}
  \begin{eqnarray}
     \ddot{y}_{\rm Dy} & = & -\omega_{\rm Dy}^2 y_{\rm Dy} -
     q_{\rm Dy} \beta \, \dot{\delta y} - q_{\rm Dy} \kappa \, \delta y \, ,\\
    \ddot{y}_{\rm K} & = & - \omega_{\rm K}^2 y_{\rm K} + q_{\rm K} {\beta} \, \dot{\delta y}  + q_{\rm K}{\kappa} \, \delta y \, ,
\end{eqnarray}
\label{eq:eqsmotion}
\end{subequations}
with $\delta y = y_{\rm Dy} - y_{\rm K}$ representing the relative displacement. The mass factors are defined as $q_{\rm Dy} = M_{r}/M_{\rm Dy}$ and $q_{\rm K} = M_{r}/M_{\rm K}$, where $M_r = M_{\rm Dy} M_{\rm K} / ( M_{\rm Dy} + M_{\rm K})$ is the reduced total mass with $M_{\rm Dy} = N_{\rm Dy} m_{\rm Dy}$ and $M_{\rm K} = N_{\rm K} m_{\rm K}$. The interaction coefficients $\beta$ and $\kappa$ represent the dissipative drag and the reactive interspecies interaction, respectively.

\begin{figure}[t]
\centering
\includegraphics[trim=0 8 0 0,clip,width=1\columnwidth]{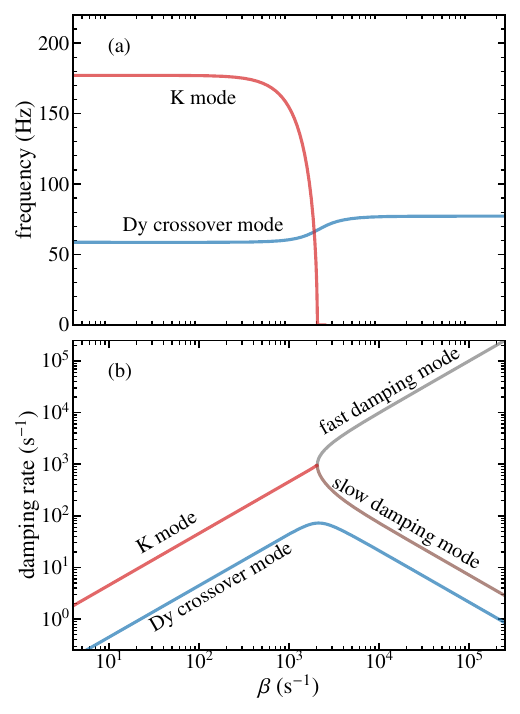}
\caption{Predictions of the theoretical model for the dependence of dipole-mode frequencies (a) and damping rates (b) on the friction parameter $\beta$. Here we assume $\kappa=0$ (no reactive coupling) and typical experimental conditions ($\omega_{\rm Dy}/2\pi = \qty{59}{Hz}$, $\omega_{\rm K}/2\pi = \qty{177}{Hz}$,
$N_{\rm Dy}/N_{\rm K} = 2.5$).}
\label{fig:TheoryCurve}
\end{figure}

\subsection{Dipole modes}

Solving the equations of motion (\ref{eq:eqsmotion}) as outlined in App.~\ref{app:eqmotion}, we obtain the complete mode spectrum with its complex-valued eigenfrequencies and eigenvectors. We find that, under {our typical experimental conditions}, the dissipative friction effect generally dominates over the reactive mean-field effect.
{Note that this stands in contrast to experiments on dipole oscillations of mixed superfluids~\cite{Ferrierbarbut2014amo, Wu2018cdo}, where the reactive coupling plays the dominant role.}
{Here we focus our discussion on the general behavior in the case of dissipative coupling, at this point setting $\kappa=0$ and deferring a discussion of the weaker effects} induced by the reactive coupling to the later Sec.~\ref{sec:Dycrossovermode}. 

{
In Fig.~\ref{fig:TheoryCurve}, we show the dipole-mode frequencies (a) and damping rates (b) as a function of the friction parameter $\beta$, assuming typical experimental conditions. 
The behavior illustrates the transition of the system from a weakly coupled regime, where the oscillations of the two species keep their individual character, to a hydrodynamic regime, where both species behave jointly.
This transition is marked by a critical value of the friction parameter ($\beta_{\rm cr} \approx 2000\,{\rm s}^{-1}$ in our example).
For weak friction, the system exhibits two damped oscillatory modes, which in the uncoupled limit correlate with the independent oscillations of the two species at their bare frequencies $\omega_{\rm Dy}$ and $\omega_{\rm K}$. With increasing dissipative coupling, the two frequencies approach each other and damping increases for both modes.
Once the critical damping rate $\beta_{\rm cr}$ is reached, one of the two oscillatory modes 
becomes  overdamped and splits up into two purely exponential modes. Under the conditions of our experiments, this splitting happens for the mode correlated with the bare oscillation of the K component and we thus refer to this mode as ``K mode''. With further increasing dissipative coupling the damping of one of these exponential modes increases (``fast damping mode''), while it decreases for the other one (``slow damping mode''). 
}

{
The second oscillatory mode, which for $\beta \rightarrow 0$ correlates with the bare Dy oscillation, exists for any strength of the dissipative coupling. In contrast to the K mode, this mode shows a smooth crossover between the regimes of weak and the strong friction, and we thus refer to it as ``Dy crossover mode''. 
For $\beta < \beta_{\rm cr}$ the coupling to the K component introduces damping along with a frequency upshift. With the friction strength reaching $\beta_{\rm cr}$, the mode exhibits maximum damping. For further increasing coupling, 
the mode damping again decreases. For $\beta \gg \beta_{\rm cr}$ the mode can be understood in terms of a locked hydrodynamic oscillation, in which both components act jointly and the system behaves like a single fluid.
}

\subsection{Case of strong friction} 
\label{ssec:strongfriction}

The case of strong friction ($\beta \gg \beta_{\rm cr}$) is of particular relevance for mixtures in the strongly interacting regime. Here one can derive rather simple analytical expressions for the frequency of the crossover mode and the damping rates of all modes. Solving the equations of motion (\ref{eq:eqsmotion}) in a perturbative way, we obtain
\begin{equation}
    \omega_{\rm c} = \sqrt{   \frac{M_1 \omega_1^2 + M_2 \omega_2^2 }{M_1 + M_2}} \, 
\label{eq:lockedfreq}
\end{equation}
for frequency of the locked hydrodynamic oscillation in the crossover mode and 
\begin{equation}
\Gamma_{\rm c} =
    \frac{\tfrac{1}{2}M_r (\omega_2^2-\omega_1^2)^2}
    {M_1 \omega_1^2 + M_2\omega_2^2} \, \beta^{-1} \, 
\end{equation}
for the corresponding damping rate.

For the slow damping mode, we obtain
\begin{equation}
\Gamma_{\rm slow} =
    \frac{(M_1+M_2) \, \omega_1^2 \omega_2^2}
    {M_1 \omega_1^2 + M_2\omega_2^2} \, \beta^{-1} \, ,
     \label{eq:Gammaslow}
\end{equation}
while for the fast damping mode the simple result is
\begin{equation}
    \Gamma_{\rm fast} =  \beta \, .
    \label{eq:Gammafast}
\end{equation}
We point out that, in the strongly interacting limit, the fast damping mode can be interpreted as a direct manifestation of the drag effect between the two species in case of a relative motion.
Measurements of damping of this particular mode, which appears on a very short time scale, allow us to extract the corresponding friction parameter $\beta$ (also refered to as ``drag coefficient'') with high accuracy.

\section{Experimental procedures}\label{sec:experimental}

\subsection{Sample preparation}\label{ssec:preparation}

The preparation of the optically trapped mixture 
has been described in our previous work \cite{Ravensbergen2018poa, Ye2022ool, Soave2023otf}. Summarized in brief, a two-species magneto-optical trap (MOT) is loaded with $^{161}$Dy from a Zeeman-slowed atomic beam and with $^{40}$K from a two-dimensional MOT. The mixture is transferred into an optical dipole trap (ODT) operating with near-infrared light (wavelength \qty{1064}{nm}). Both species are spin polarized in their lowest hyperfine sublevels $\ket{F,m_F}=\ket{21/2,-21/2}$ and $\ket{9/2,-9/2}$, respectively. Forced evaporative cooling, based on universal dipolar collisions within the Dy component \cite{Lu2012qdd, Aikawa2014rfd} and sympathetic cooling of K by Dy \cite{Ravensbergen2018poa}, brings the sample into degeneracy. Evaporative cooling is most efficiently performed at a low magnetic field of \qty{250}{mG}. The sample is then 
transferred to higher magnetic fields close to the interspecies Feshbach resonance near \qty{7.3}{G} \cite{Ye2022ool, Soave2023otf}, which we use for interaction tuning.

The final ODT, in which all experiments are carried out, is realized with trap light at \qty{1547}{nm}. This is further in the infrared than the trap light used in all our previous experiments on Dy-K mixtures. We found that this reduces inelastic trap losses resulting from coupling to excited molecular states \cite{Soave2023otf}. As a consequence of this change, the ratio of optical polarizabilities and thus trap depths for both species \cite{Ravensbergen2018ado} is different from our previous work. With scalar polarizabilities of {383}\,a.u.\ for K \cite{UDportal} and {173}\,a.u.\ estimated for Dy \cite{Dzuba2011dpa, Li2017oto} we obtain a trap frequency ratio of $\omega_{\rm K}/\omega_{\rm Dy} = 2.99(4)$. We employ a standard crossed-beam ODT, realized with a
horizontal beam (power \qty{1.5}{W}, waist \qty{87}{\micro\meter}) combined with a vertical beam (power \qty{310}{\milli\watt}, waist of \qty{123}{\micro\meter}). The trap provides a nearly axially symmetric optical potential with radial frequencies $\omega_{\rm Dy}/2\pi = \qty{59}{Hz}$ and $\omega_{\rm K}/2\pi = \qty{177}{Hz}$ (relevant for the dipole oscillations), and axial frequencies of about four times less. This corresponds to mean trap frequencies of $\bar{\omega}_{\rm Dy}/2\pi = \qty{37}{Hz}$ and $\bar{\omega}_{\rm K}/2\pi = \qty{112}{Hz}$.

In our trap, we reach typical conditions of $N_{\rm Dy} = 5\times10^4$ Dy atoms and $N_{\rm K} = 2\times10^4$ K atoms a temperature of $T = \qty{100}{nK}$. With Fermi temperatures 
$T_F^{\rm Dy} = \sqrt[3]{6N_{\rm Dy}} \, \hbar \bar{\omega}_{\rm Dy} / k_B
\approx \qty{120}{nK}$ and 
$T_F^{\rm K} = \sqrt[3]{6N_{\rm K}} \, \hbar \bar{\omega}_{\rm K} / k_B \approx \qty{260}{nK}$, we are in a regime of moderate Fermi degeneracy~\footnote{Our temperatures are currently limited by residual heating effects in the transfer from the low evaporation magnetic field to the magnetic field region of interest, presumably by crossing many narrow intraspecies Dy Feshbach resonances \cite{Soave2022lff}.
}.
As discussed in App.~\ref{app:overlap}, the spatial profiles of the two components nearly match (the Dy cloud being slightly larger), which provides us with a good overlap between both species.

The atom number ratio $N_{\rm Dy}/N_{\rm K}$ can be varied by changing the loading times of the MOTs. We have investigated a maximum range between $0.5$ and $7.5$, but for most of our experiments we found optimum conditions for typical ratios 
near 2.5 (Sec.~\ref{sec:Dycrossovermode}) or 4.5 (Sec.~\ref{sec:dampingmodes}).

Magnetic levitation \cite{Weber2003bec, Ravensbergen2018poa} plays a very important role in our experiments. Without a vertical magnetic field gradient, which (partially) compensates for the effect of gravity, our rather shallow optical trap would not be able to keep the atoms. A `magic' levitation gradient of \qty{2.63}{G/cm}, with equal gravitational sag of both species \cite{Lous2017toa, Soave2023otf}, ensures maximum spatial overlap between the two components. A controlled deviation from this particular gradient induces a vertical displacement between the two components. Correspondingly, we apply short gradient pulses (or a sequence of such pulses) as versatile tools to excite the two-species dipole modes. 


\begin{figure*}[htb]
\centering
\includegraphics[trim=5 5 5 0,clip,width=2\columnwidth]{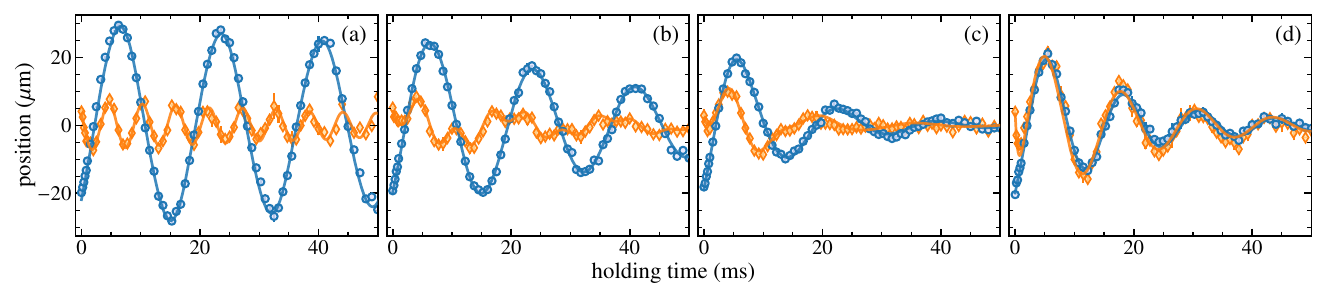}
\caption{Center-of-mass oscillations of the Dy (blue) and K (orange) clouds for increasing interaction strength. The vertical position is obtained from time-of-flight images and plotted versus the hold time in the trap. The atom number ratio is fixed, $N_{\rm Dy}/N_{\rm K}=2.42$. (a) Near the zero crossing of the scattering length, where $x\rightarrow \pm\infty$, the two clouds oscillate independently. (b) and (c) When approaching the resonance, $x = 1.6$ and $0.7$, increasing interaction effects are observed. (d)~On resonance, $x=0$, the interspecies interaction results in a locked hydrodynamic motion of both components. The positions of the clouds are detected after a time of flight of \qty{10}{ms} for Dy and \qty{5}{ms} for K. For better comparison, the K signal is rescaled by a factor of 2. The uncertainties are smaller than the size of the symbols and the solid curves represent fits based on our theoretical model (see text). }
\label{fig:Oscillation}
\end{figure*}

\subsection{Interaction tuning}\label{ssec:tuning}

We control the interacting strength in the mixture via a magnetically tuned Feshbach resonance \cite{Ye2022ool, Soave2023otf}, which is centered at $B_0 = \qty{7.276(2)}{G}$. In the resonance region, well-isolated from other resonances, the interspecies $s$-wave scattering length is given by
\begin{equation}
a= a_{\rm bg} - \frac{A}{\delta B} \, a_0 ,
\label{eq:scattlength}
\end{equation}
where $\delta B = B-B_0$ represents the magnetic detuning, the parameter $A = \qty{24.0(6)}{G}$ characterizes the strength of the resonance, 
and $a_0$ denotes the Bohr radius. The quantity $a_{\rm bg} = +23(5)\,a_0$ represents the background scattering length, which is rather small and can be neglected close to the resonance center.

For convenience in the following discussion, we introduce the dimensionless interaction parameter 
\begin{equation}
x \equiv 1000\,a_0/a \, , 
\label{eq:definex}
\end{equation}
which close to resonance is proportional to the magnetic detuning, $x \approx -\delta B/ \qty{24}{mG}$. The root-mean-square fluctuations of the magnetic field stay within typically \qty{2.5}{mG}, which translates into an interaction parameter uncertainty of $\sigma_x \approx 0.1$.

\subsection{Excitation and detection}\label{ssec:excitedetect}

To excite the various two-species dipole modes we apply pulsed changes of the magnetic levitation gradient, thus inducing motion in the vertical direction. These pulses are always applied at a large magnetic detuning 
where the interspecies interaction 
is negligible on the timescale of the excitation process. In all present experiments, we choose {$\delta B = -132$\,mG}, corresponding to $x=+4.9$.
Having induced the motion of the two clouds in the weakly interacting regime, the magnetic field is ramped within typically \qty{0.2}{ms} to the target value where the mode is studied at a well-defined interaction strength. This ramp, starting from below the resonance, always increases the magnetic field strength, which is crucial to avoid molecule association \cite{Chin2010fri, Soave2023otf}.

Our typical magnetic gradient pulses are applied for \qty{0.2}{ms} and increase the levitation gradient from its magic value (\qty{2.63}{G/cm}) to values about \qty{1}{G/cm} higher. The higher gradient pulls both species out of the equilibrium position. This predominantly affects the Dy atoms because of their exceptionally large magnetic dipole moment of $10\,\mu_B$ (as compared to $1\,\mu_B$ for K); here $\mu_B$ is Bohr's magneton.
To optimize the excitation of particular modes we can apply a second gradient pulse, which depending on the delay cancels or enhances the excited motion in a species-selective way. Another important parameter for the optimization of the excitation sequence is the time delay after which the magnetic field is ramped to its target value, which controls the point where the system is converted from the regime of independently oscillating motions to a coupled motion of both components. The details of the procedure depend on the particular mode to be excited. 

After a variable hold time, in which the dynamics of interest takes place, we turn off the ODT and quickly switch the magnetic field to a large detuning ($\delta B = +472\,$mG, corresponding to $x = -36$). At this field, the two components expand ballistically without any significant further interaction. We finally apply standard time-of-flight imaging after 10\,ms (for Dy) or 5\,ms (for K) of free expansion, from which we determine the vertical center-of-mass positions of the two species.

Inelastic losses, reducing the number of trapped atoms during the hold time in the trap, generally stay well below a few 10\% and have a weak effect on the oscillation curves. Only in a narrow range (few mG wide) very close to the center of the resonance, we have observed faster losses. In the worst case, we have measured lifetimes in the mixture of about 150\,ms for Dy and 50\,ms for the K component. 

Typical oscillation signals are shown in Fig.~\ref{fig:Oscillation} for increasing strength of the interspecies interaction. The observed behavior follows the basic predictions of the model introduced in Sec.~\ref{sec:modespectrum}. For the non-interacting case (a), both species oscillate independently at their individual trap frequencies. In this collisionless regime, damping due to interspecies interaction is essentially absent. Weak residual damping may be attributed to trap imperfections, such as anharmonicities.
When approaching the resonance, the observed oscillations in (b) and (c) reveal increasing interaction effects. While both components show collisional damping, the K signal also reveals the emergence of two frequency components, which we interpret as a superposition of the K mode with the Dy crossover mode. In our experiments, the K component generally shows much stronger interaction effects because of the much larger total mass of the Dy cloud ($161 N_{\rm Dy}/40 N_{\rm K}\approx 10$). For the strongest interaction, realized on top of the resonance (d), we observe the expected locked hydrodynamic oscillation, in which both species oscillate jointly together with a well-defined frequency according to Eq.~(\ref{eq:lockedfreq}). In the regime of strong interactions, damping becomes weaker with increasing interaction, which confirms a central prediction pointed out in Sec.~\ref{sec:modespectrum} for the Dy crossover mode. 

The on-resonance oscillation curves in Fig.~\ref{fig:Oscillation}(d) reveal another interesting feature. The two curves overlap almost perfectly, owing to the locked hydrodynamic oscillation of the two species in the Dy crossover mode, except for an initial transient in the first millisecond. This is a signature of the fast damping mode, as we will discuss in detail in Sec.~\ref{sec:dampingmodes}.

\subsection{Fit analysis of oscillation curves}\label{ssec:fitting}

The frequencies and damping rates of the dipole modes at a specific interaction strength can be extracted by fitting the time-dependent solutions $y_{\rm Dy}(t)$ and $y_{\rm K}(t)$ of our model 
to the observed center-of-mass motion of the two species. See, for example, the solid lines in Fig.~\ref{fig:Oscillation}. 
For an oscillating mode, the in-trap positions, which describe the motion before time-of-flight expansion, follow damped harmonic oscillations
\begin{equation}
{y}_i^{\rm trap}(t) = \hat{A}_{i} \, e^{-\gamma t}\cos(\omega t + \phi_{i})  \, ,
\end{equation}
where $i$ = Dy, K. While, for a particular mode, the two species share the same frequency $\omega$ and the same damping rate $\gamma$, the amplitudes and phases are generally different.
Taking into account the ballistic motion of the clouds during the free time-of-flight expansion, the resulting positions are described by
\begin{equation}
\begin{aligned}
{y}_i(t) = \hat{A}_{i} \, e^{-\gamma t} \,
[&
(1 - \gamma t_{\rm TOF}) \cos(\omega t + \phi_{i})  
\\& - \omega t_{\rm TOF} \sin(\omega t + \phi_{i})
] \, .
\end{aligned}
\label{eq:oscmodefit}
\end{equation}
For a damping mode ($\omega=0$), the behavior simplifies to
\begin{equation}
{y}_i(t) = \hat{A}_{i} \, e^{-\gamma t} \,
(1 - \gamma t_{\rm TOF})
 \, .
\label{eq:dampmodefit}
\end{equation}

Below the critical damping point, we fit the observed behavior by assuming a superposition of two oscillating modes according to Eq.~(\ref{eq:oscmodefit}). For each mode, there are six free parameters: frequency $\omega$, damping rate $\gamma$, and the species-dependent  amplitudes $A_{\rm Dy}$ and $A_{\rm K}$ and phases $\phi_{\rm Dy}$ and $\phi_{\rm K}$. In total we have to deal with $14$ free parameters, six per mode and two additional parameters for the equilibrium positions of the two species without excitation.
This complexity requires combined (not individual) fits to the data recorded for $y_{\rm Dy}(t)$ and $y_{\rm K}(t)$. Above the critical damping point the complexity of the fit remains the same as the fit has to take into account one oscillatory mode and two damping modes, the latter according to Eq.~(\ref{eq:dampmodefit}). 

Finally, to fully characterize each mode, the quantities of interest are its frequency $\omega$ and damping rate $\gamma$ (corresponding to the complex-valued eigenfrequencies) and the relative amplitude $A_{\rm K}/A_{\rm Dy}$ and phase $\Delta\phi \equiv \phi_{\rm K}-\phi_{\rm Dy}$ (representing the eigenvectors). A damping mode is fully characterized by the damping rate $\gamma$ and the amplitude ratio $A_{\rm K}/A_{\rm Dy}$. In the experiments described in the following section, we measure all these quantities in the full range of interaction strengths across the resonance.

\section{Experimental results}\label{sec:results}

In this Section, we present our main experimental results on the two-species dipole mode spectrum across resonance. In Sec.~\ref{sec:overview}, we first give an overview of the observed spectrum. Then, in Sec.~\ref{sec:Dycrossovermode}, we discuss the Dy crossover mode in detail. This particular mode exists for any interaction strength and allows for an accurate determination of the parameters of our interaction model.
Then, in Sec.~\ref{sec:dampingmodes}, we turn our attention to the damping modes.

\subsection{Mode spectrum overview} 
\label{sec:overview}

\begin{figure}
\centering
\includegraphics[trim=0 6 0 0,clip,width=1\columnwidth]{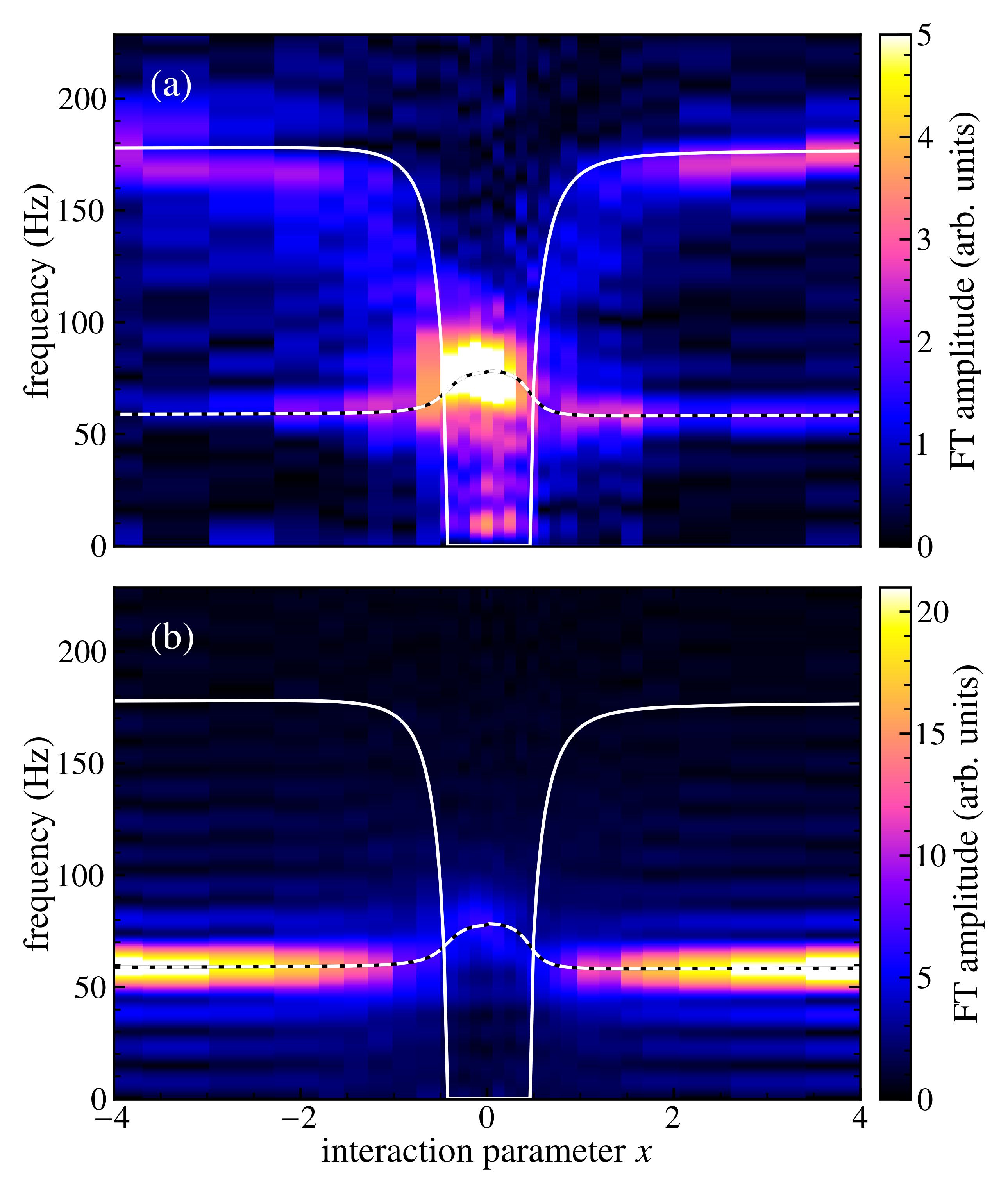}
\caption{Mode spectrum across the resonance. The pseudo-color plot shows the amplitude spectrum derived by Fourier transforming the oscillation signals recorded for (a) the K atoms and for (b) the Dy atoms. The solid white line and the black-white dashed line are the theoretical curves for the K mode and the Dy crossover mode (for details see Sec.~\ref{sec:Dycrossovermode}). Note that the horizontal stripes visible in (b) are an artefact related to the Fourier transform of the rectangular time window.
}
\label{fig:FFTsignal}
\end{figure}

We recorded a large set of oscillation curves for typical atom numbers 
$N_{\rm Dy} = 5.8 \times10^4$,
$N_{\rm K} = 2.4 \times10^4$ 
($N_{\rm Dy}/N_{\rm K} = 2.42$)
and temperatures $T \approx 100\,$nK.
The 30 recorded curves (per species) cover an interaction parameter range of $-6 \le x \le +6$ ($|a| \ge 167\,a_0$). Each curve consists of 90 data points with 3 to 5 repetitions taken in a time interval of \SI{70}{ms} after the excitation by a double-pulse sequence. To get a first impression of the mode spectrum, we carried out a fast Fourier transform (FFT) for each curve and visualize the resonance behavior of the amplitude spectrum in the pseudo-color plots of Fig.~\ref{fig:FFTsignal}, where panel (a) refers to the signal obtained from the K component and (b) from the Dy component.

Already the simple FFT analysis demonstrates the essential features of the mode spectrum across resonance. In the K signal (a), we can identify both modes with their bare frequencies of 177\,Hz and 59\,Hz in the weakly interacting regime. 
On resonance, only the locked oscillation ($\sim$78\,Hz) is observed. In the Dy signal (b), the off-resonant K mode excitation is too weak to be observed in the data, but the resonant crossover to the locked mode can be clearly seen.
For comparison,
we also show theoretical curves for the frequencies of the Dy crossover mode (dashed line) and the K mode (solid line) as derived from our model and explained in detail in the following Sections.

\subsection{Crossover mode}
\label{sec:Dycrossovermode}

\begin{figure}[!tb]
\centering
\includegraphics[trim=0 6 0 0,clip,width=1\columnwidth]{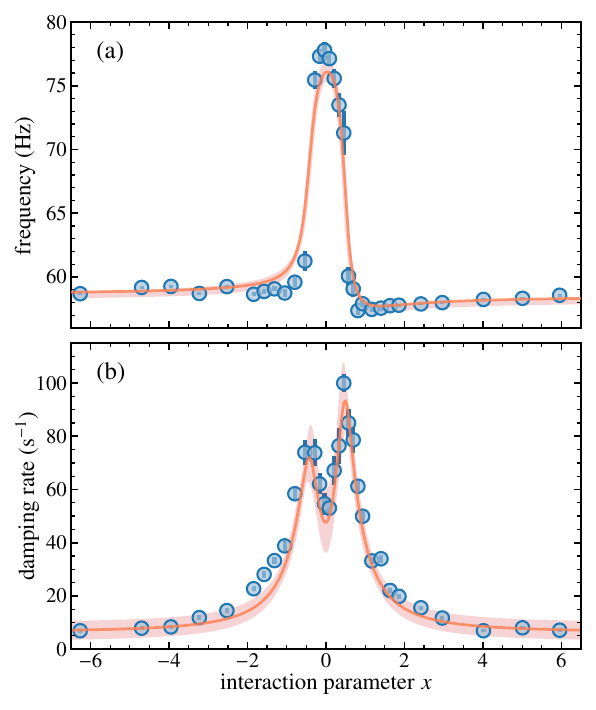}
\caption{Resonance behavior of (a) frequency and (b) damping rate of the Dy crossover mode. The blue data points represent the measured values with 1$\sigma$ error bars (in most cases smaller than the symbol size). The red solid curves show the results of a joint fit to frequency and damping based on the model presented in Sec.~\ref{sec:modespectrum}. 
The red shaded area represents the 95\% confidence band of the fit.
}
\label{fig:DymodeFreqAndDamping}
\end{figure}

\begin{figure}[!tb]
\centering
\includegraphics[trim=0 6 0 0,clip,width=1\columnwidth]{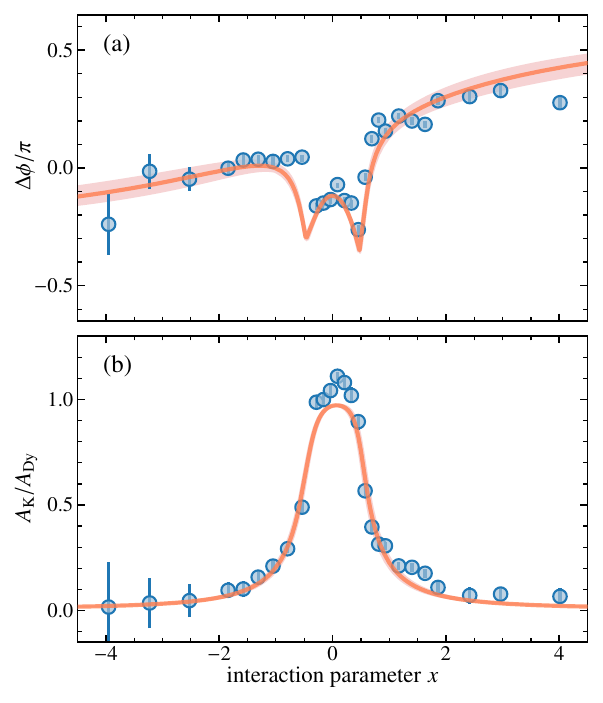}
\caption{(a) Phase shift $\Delta\phi=\phi_{\rm K}-\phi_{\rm Dy}$ and (b) amplitude ratio $A_{\rm K}/A_{\rm Dy}$ of the two species oscillating in the Dy crossover mode. The blue data points represent the measured values with 1$\sigma$ error bars. The red solid curves show the result of our model with the same parameter values as used in  Fig.~\ref{fig:DymodeFreqAndDamping} (no independent fit carried our here).  }
\label{fig:PhaseAndAmpRatio}
\end{figure}

We analyze the recorded oscillation curves as explained in Sec.~\ref{ssec:fitting}, extracting full information on the mode spectrum. Based on the theoretical model of Sec.~\ref{sec:modespectrum} and model assumptions for the resonance dependence of the interaction coefficients $\beta(x)$ and $\kappa(x)$, we can then fit the full resonance behavior. Here, as a case of particular interest, we focus on the resonance behavior of the Dy crossover mode. 

\subsubsection{Frequency and damping rate}

The data points in Fig.~\ref{fig:DymodeFreqAndDamping} 
show the experimental values obtained for the frequency and damping rate. 
In panel (a), the frequency change from the bare oscillation (56\,Hz) for weak interaction to the locked oscillation (78\,Hz) on resonance is clearly visible. The locked oscillation frequency is in full agreement with Eq.~(\ref{eq:lockedfreq}).
The damping rate measurements in panel (b) reveal maxima at $|x| \approx 0.5$ ($|a| \approx 2000\,a_0$), 
indicating that the critical damping rate $\beta_{\rm cr}$ is reached at the corresponding interaction strength. Closer to resonance, a clear decrease of damping is observed, in accordance with the behavior in Fig.~\ref{fig:Oscillation}(d).

While all these observations can essentially be understood within our model as a result of friction (dissipative coupling) between the two species, our data also reveal a distinct asymmetry, 
which manifests itself in slightly  higher frequencies measured on the side of attractive interaction ($x<0$) and somewhat larger damping rates observed on the repulsive side ($x>0$). This asymmetry points to the effect of reactive interaction, which we have so far neglected in our discussion of the mode spectrum in Sec.~\ref{sec:modespectrum} by setting 
$\kappa = 0$. 

The phase difference and the relative amplitude of the two-species oscillations provide additional information on the resonance behavior of the mode. 
Our corresponding results are displayed
in Fig.~\ref{fig:PhaseAndAmpRatio}. 
The phase in panel (a) exhibits a pronounced asymmetry, which turns out to be the most pronounced manifestation of the reactive interaction between the two components.
The amplitude ratio in panel (b) clearly shows that, in the weakly interacting regime, the mode is dominated by the oscillation of the Dy component with a small admixture of K oscillation. For increasing interaction this smoothly connects to a locked oscillation, where both components feature essentially the same amplitude.

\subsubsection{Fit analysis}
\label{sssec:fitanalysis}

For a quantitative interpretation of the experimental observations in terms of the model presented in Sec.~\ref{sec:modespectrum}, we introduce the functions $\beta(x)$ and $\kappa(x)$ to describe the dissipative and reactive coupling across the resonance. Based on standard theory of resonant $s$-wave scattering and adopting a collisional mean-field approach (see App.~\ref{app:cloudinteraction}), we write
\begin{equation}
    \beta(x) = \frac{\beta^*}{x^2 + x_c^2} \, 
    \label{eq:betax}
\end{equation}
and
\begin{equation}
    \kappa(x) = -\frac{\kappa^* x}{x^2 + x_c^2} \, .
        \label{eq:kappax}
\end{equation} 
Here the quantities $\beta^*$ and $\kappa^*$ characterize the overall strength of the dissipative and the reactive coupling, respectively, depending on the  particular experimental conditions. The parameter $x_c = 1000\,a_0/a_c$ corresponds to a characteristic value of the scattering length, $a_c$, above which the interaction is limited by the finite collisional momenta. In a thermal gas, this length is determined by the thermal de~Broglie wavelength of the relative motion, $a_c \approx 1.21\,{\lambda_{\rm th}}/(2 \pi)$ with $\lambda_{\rm th} = \sqrt{2 \pi \hbar^2/(m_r k_B T)}$; see App.~\ref{app:beta}.

We perform a combined fit to the complete set of measurements of frequency and damping rate, as displayed in Figs.~\ref{fig:DymodeFreqAndDamping}(a) and (b). The three main parameters of the fit are $\beta^*$, $\kappa^*$ and $a_c$. 
The bare K frequency is fixed to a value $\omega_{\rm K}/2\pi = 177.2\,$Hz, accurately determined in a separate measurement. The bare Dy frequency, {for which separate measurements are subject to relatively large uncertainties~\footnote{The bare oscillation frequencies can be measured with standard methods \cite{Ravensbergen2018ado}. For both species, the absolute uncertainties are similar, which means that typical uncertainties are three times large for Dy than for K. Moreover, the trap anharmonicity has a stronger effect in the shallower potential experienced by the Dy atoms.},} is kept as a free parameter and its value is extracted from the fit ($\omega_{\rm Dy}/2\pi = 58.6\,$Hz). 
In the fit to the damping rate data, we phenomenologically take into account a weak background ($\sim 6$\,s$^{-1}$), which was measured in the absence of interspecies interactions near the zero crossing of the scattering length.

Our fit (solid lines in Fig.~\ref{fig:DymodeFreqAndDamping}) describes the experimentally observed resonance behavior remarkably well. All basic features of the Dy crossover mode are reproduced and satisfying quantitative agreement is reached within the uncertainties. For the dissipative coupling strength we obtain the parameter value  $\beta^* =  750(60)\,\mathrm{s}^{-1}$, which is fully consistent with the theoretical result of $680(120)\,\mathrm{s}^{-1}$  from the collisional model described in App.~\ref{app:cloudinteraction}. Maximum damping is observed at $|x| = 0.46$ ($|a|= 2200\,a_0$), which according to Eq.~(\ref{eq:betax}) corresponds to a critical value of $\beta_{\rm cr} = 2100\,\textrm{s}^{-1}$ for the friction parameter. This fully agrees with the expectation from Fig.~\ref{fig:TheoryCurve}. The fit also provides us with a value for the friction parameter on resonance, $\beta_0 = \beta^*/x_c^2 \approx 5000\,\textrm{s}^{-1}$. 

For the finite-momentum limitation of the interaction strength, as imposed by unitarity, our fit yields $x_c = 0.38(3)$, corresponding to $a_c = 2600(200)\,a_0$. This is essentially consistent with the value $a_c = 3500\,a_0$ as expected under the present experimental conditions for collisions in a thermal gas; see App.~\ref{app:beta}. In view of the moderate Fermi degeneracy in our mixture, it is also close to the
 length scales set by the inverse Fermi wavenumbers of the two species, $1/k_F^{\rm Dy} \approx 2100 \,a_0$ and $1/k_F^{\rm K} \approx 2900 \,a_0$.

Remarkably, the fit also reproduces the slight asymmetry in the data of Fig.~\ref{fig:DymodeFreqAndDamping}, for which it yields a value of $\kappa^* = 7.7(1.5)\times 10^4\,\textrm{s}^{-2}$ for the reactive coupling strength. This is about twice larger than the result of $\sim\,4 \times 10^4\,\textrm{s}^{-2}$ calculated within our mean-field approach in App.~\ref{app:cloudinteraction}, but still reasonable in view of the simplifying assumptions of the theoretical approach, such as neglecting interaction-induced distortions of the spatial cloud profiles.

Keeping the parameter values as extracted from the fit above, we also apply our model to describe the resonance behavior of the phase difference 
$\Delta\phi \equiv \phi_{\rm K}-\phi_{\rm Dy}$ 
and the amplitude ratio $A_{\rm K}/A_{\rm Dy}$  between the two components (red solid lines in Fig.~\ref{fig:PhaseAndAmpRatio}). Without any further fitting, we find excellent agreement with the measured values (blue data points). This  again confirms the validity of the model introduced in Sec.~\ref{sec:modespectrum}.

{A particularly interesting observable is the phase difference $\Delta \phi$ between the oscillations of the Dy and the K component in the crossover mode.
Here the contribution by the reactive mean-field effect plays an important role over the whole range of interaction investigated. The observed behavior of the phase can only be understood by taking into account the reactive term. For $\kappa=0$, our model would predict a flat phase across resonance outside of the locking region, whereas a finite value of $\kappa$ breaks the symmetry and shifts the phase  even for moderate interaction strength in the non-resonant regime ($|x|>1$). Note that, away from resonance, the strength of the reactive coupling effect falls of proportional to $x^{-1}$, whereas the dissipative effect drops faster with $x^{-2}$.
In particular the larger phase difference on the repulsive side can be understood as a result of a delay originating from the repulsion between the two clouds.}

Our results provide a complete characterization of the two-species `crossover' dipole mode that exists for any interaction strength across a resonance, with full information on the complex-valued eigenfrequency (frequency and damping rate) and on the eigenvector (relative phase and amplitude ratio). 
While previous experiments \cite{Gensemer2001tfc, Ferlaino2003doi, Delehaye2015cva, Finelli2024ula} have demonstrated the
effect of a joint oscillation of both species as an effect of dissipative coupling (interspecies friction), 
we also observe the weaker additional effect of reactive coupling, caused by attraction or repulsion between the two components.

\subsubsection{Dependence on atom number ratio}

\begin{figure}
\centering
\includegraphics[trim=0 6 0 0,clip,width=1\columnwidth]{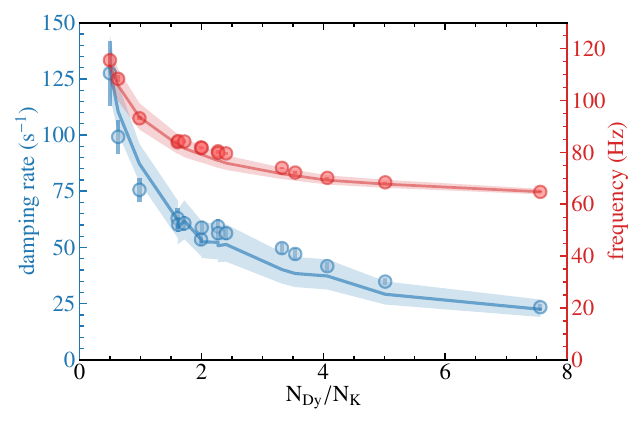}
\caption{Frequency and damping rate of the Dy crossover mode on resonance ($x=0$) as a function of the atom number ratio. With increasing $N_{\rm Dy}/N_{\rm K}$, the temperature decreases from 170 to \SI{130}{nK}. The bullet symbols represent the experimental data with 1$\sigma$ uncertainties. The solid curves show the theoretical behavior according to our model (see text). The shaded regions represent the effect of $\pm10\%$ atom number uncertainties.}
\label{fig:FreqAndDampingwithNr}
\end{figure}

To further investigate the validity of the theoretical model, we examine the frequency and the damping rate of the Dy crossover mode on resonance in a wide range of atom number ratios $N_{\rm Dy}/N_{\rm K}$ from 0.5 to 7.5. In Fig.~\ref{fig:FreqAndDampingwithNr}, we show the experimental results (data points) in comparison with the predictions of the theoretical model (solid lines). 
In this set of measurements, the particular experimental conditions (absolute atom numbers $N_{\rm Dy}$, $N_{\rm K}$, and temperature $T$) vary, depending on the particular loading sequence applied. 
{For the lowest number ratio, 
$N_{\rm Dy} = 4.2\times10^{4}$, $N_{\rm K} = 8.3\times 10^4$, and $T = 170\,$nK,
For the highest ratio, 
$N_{\rm Dy} = 1.5\times10^5$, $N_{\rm K} = 2.0\times10^4$, and $T = 130\,$nK.}

For calculating the theoretical frequency and damping rate values, we have to consider the individual conditions for each data point.
We separately measure the numbers $N_{\rm Dy}$, $N_{\rm K}$ and the temperature $T$. Following App.~\ref{app:beta}, we calculate the value of the friction parameter $\beta$. The reactive interaction can be neglected on resonance ($\kappa =0$). Again applying the model of Sec.~\ref{sec:modespectrum} we can finally calculate the mode frequency and damping rate.

With increasing $N_{\rm Dy}/N_{\rm K}$, the measurements demonstrate the behavior of the oscillation frequency expected from  Eq.~(\ref{eq:lockedfreq}): The frequency of the locked oscillation changes from a value near the bare K frequency to a value close to the bare Dy frequency. On the other hand, the damping rate decreases with increasing $N_{\rm Dy}/N_{\rm K}$, which reflects the decreasing effect of the K component on the oscillation of Dy, the latter having a much larger total mass.

\subsection{Damping modes}
\label{sec:dampingmodes}

In this Section, we focus on the two non-oscillating damping modes. According to our model, introduced in Sec.~\ref{sec:modespectrum}, the fast and the slow damping mode exist if the friction parameter exceeds a critical value ($\beta > \beta_{\rm cr}$). These two modes coexist with the oscillating crossover mode discussed before.

The experiments have been carried out in the same optical trap setting as in Sec.~\ref{sec:Dycrossovermode} ($\omega_{\rm Dy}/2\pi = 58.6$\,Hz, $\omega_{\rm K}/2\pi = 177.2$\,Hz), but with a larger number of initially loaded Dy atoms. After evaporative cooling, we keep
$N_{\rm Dy} = 8.6 \times 10^4$ and $N_{\rm K} = 1.9 \times 10^4$ atoms, corresponding to a number ratio of $N_{\rm Dy}/N_{\rm K}=4.5$. The temperature of the mixture is $T = 130$\,nK. {Our choice of a larger number of Dy atoms (as compared to the experiments on the crossover mode) was originally motivated by the idea to realize a broader magnetic-field range where hydrodynamic behavior occurs and thus to obtain data with better resolution on the interaction parameter scale. However, it turned out that this effect is rather small and that the overall damping behavior is quite robust against changes in the number ratio.}

For excitation and detection, we employ the methods described in Sec.~\ref{ssec:excitedetect}. The mixture is first kicked by a single magnetic-gradient pulse under conditions of negligible interspecies interaction. 
After a carefully chosen time delay, the interspecies interaction is introduced by rapidly ramping the magnetic field to its target value. For exciting the fast damping mode, the delay is chosen to maximize the velocity difference between the two species, while minimizing the spatial displacement. For the slow damping mode, the delay is set to maximize the initial displacement, keeping the velocity difference small.
For detection, we perform standard time-of-flight imaging, {extracting the center-of-mass velocities at the time of release from the corresponding displacements observed after long times of ballistic flights ($t_{\rm TOF} = 10$\,ms for Dy, and 5\,ms for K).}
In addition, for the slow mode, {we also apply} {\it in-situ} imaging to detect the center-of-mass positions of the two components.

\begin{figure}
\centering
\includegraphics[trim=0 0 0 0,clip,width=1\columnwidth]{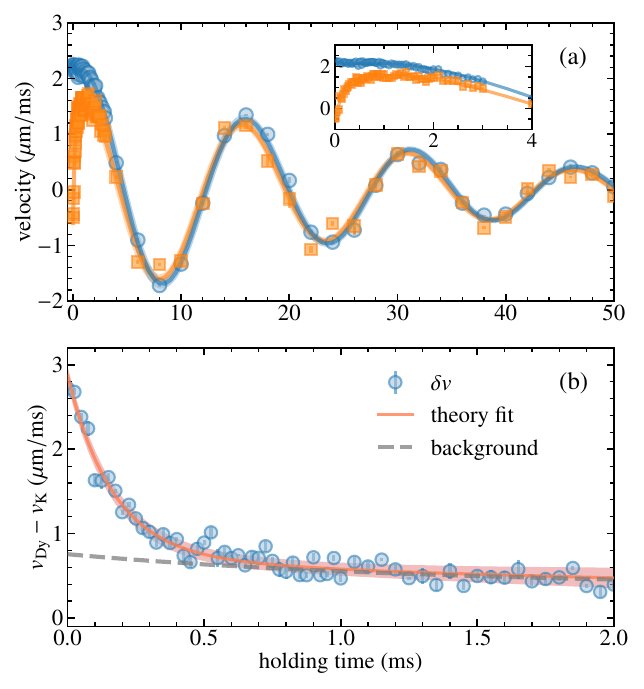}
\caption{Fast damping mode on resonance for $N_{\rm Dy}/N_{\rm K}=4.5$. In (a), we show the center-of-mass velocities of the two components, Dy (blue) and K (orange), as a function of the hold time in the ODT. The solid curves represent fits based on a superposition of the two damping modes and the crossover mode, as detailed in Sec.~\ref{ssec:fitting}. The inset zooms into the initial behavior from 0 to 4 ms. In (b), we plot the velocity difference between Dy and K in the first 2 ms. While the solid curve represents the result of the full fit, the dashed line illustrates  the slowly varying background resulting from both the slow mode and the crossover mode. The difference between the solid and the dashed curve corresponds to the fast mode. The symbols represent the experimental data with standard errors derived from five repetitions per hold time. The shaded region indicates the 95\% confidence band of the fit.}
\label{fig:FastModeDemo}
\end{figure}

\subsubsection{Fast damping mode}

The fast damping mode  
appears as a direct manifestation of the rapid reduction in the relative velocity of the two species caused by strong friction in the resonantly interacting regime. 
Figure~\ref{fig:FastModeDemo}(a) demonstrates the basic behavior as observed for resonant interspecies interaction ($x=0$). 
{Right after the dipole-mode excitation in the weakly interacting regime and a rapid quench onto resonance, which effectively takes about $20\,\mu$s~\footnote{During our 200-$\mu$s linear ramp from the preparation field ($\delta = -132\,$mG) onto resonance, the systems spends only about $20\,\mu$s in the strongly interacting regime. Thus, the interaction quench is almost an order of magnitude faster than the fastest response of the system.}, the initial velocities (at zero hold time) of the K and Dy clouds differ substantially.} Then, within less than one millisecond, the system reaches the regime of locked oscillations (inset).
Our fitting model (solid lines) reproduces the observed behavior very well, including the initial transient. Figure~\ref{fig:FastModeDemo}(b) demonstrates the evolution of the velocity difference with time.
The fast mode shows the expected exponential decay with a short damping time of about $1/\Gamma_{\rm fast} = 170\,\mu$s. {It appears on a slowly varying background, which is caused} by the slow damping mode together with the oscillating crossover mode.

The fast damping also highlights the very large rate of elastic collisions in the resonantly interacting mixture. With the approximation $\beta \approx \Gamma_{\rm fast}$, see Eq.~(\ref{eq:Gammafast}), we obtain $\Gamma_{\rm coll} = 2.0 \times 10^8\,{\rm s}^{-1}$ for the total rate of elastic collisions, see App.~\ref{app:beta} for details. This corresponds to mean rates per Dy atom and per K atom as large as
$\Gamma_{\rm coll}/N_{\rm Dy} = 2\,300\,{\rm s}^{-1}$
and
$\Gamma_{\rm coll}/N_{\rm K} = 10\,500\,{\rm s}^{-1}$,
respectively.

\begin{figure}
\centering
\includegraphics[trim=0 0 0 0,clip,width=1\columnwidth]{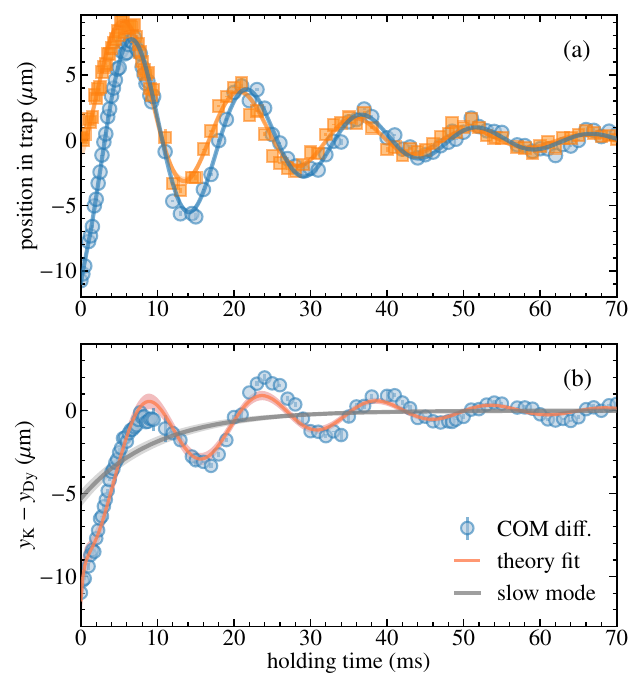}
\caption{Slow damping mode on resonance for $N_{\rm Dy}/N_{\rm K}=4.5$. (a) Temporal evolution of the COM positions of the Dy (blue) and K (orange) clouds in the ODT. The solid curves are the fits based on our theoretical model and procedures outlined in Sec.~\ref{ssec:fitting} ($t_{\rm TOF}=0$ for {\it in-situ} imaging). (b) Difference between COM positions for the Dy and K clouds. The red solid curve corresponds to the full theoretical model with all three modes, and the gray solid curve shows the contribution of the slow damping mode. The blue symbols represent the experimental data point with standard errors obtained from five repetitions per point. The shaded region around the fit curves indicates the 95\% confidence band.}
\label{fig:SLowModeDemo}
\end{figure}

\subsubsection{Slow damping mode}

The slow damping mode corresponds to an overdamped motion, in which the spatially displaced clouds relax slowly towards the equilibrium position in the trap center.
In Fig.~\ref{fig:SLowModeDemo}, we show an example, which was recorded on resonance by {\it in situ} imaging. Here the initial position difference is about \SI{10}{\micro m}, with the K cloud being nearly in the trap center, while the Dy cloud is displaced. {In panel (a), the center-of-mass positions of both components show damped harmonic oscillations, which result from an excitation of the crossover mode (discussed before in Sec.~\ref{sec:Dycrossovermode}). In the position difference, as shown in panel (b), the oscillations are reduced and an exponentially decaying background shows up clearly. The latter corresponds to the slow damping mode.}

A fit based on our theoretical model (solid lines in Fig.~\ref{fig:SLowModeDemo}) reproduces the observed behavior and yields a value of 130\,s$^{-1}$ for the damping rate $\Gamma_{\rm slow}$ on resonance. However, from corresponding measurements using the time-of-flight method (not shown), we obtain a somewhat higher value of 190\,s$^{-1}$, which is not consistent with the above value within the statistical uncertainties. A closer inspection of the fitting analysis reveals an apparent effect of systematic differences between the observed behavior and the fitting model. Although these differences remain rather small, we find that the best estimates obtained for the values of the fit parameters can significantly depend on the initial conditions and the time window selected for the fits. This problem does not occur for all other modes, where the fitting turns out to be very robust. Nevertheless, our results clearly confirm the existence of the slow damping mode on resonance and show that it is damped with a rate between the crossover mode ($\Gamma_{\rm c} = 35\,{\rm s}^{-1}$) and the fast damping mode ($\Gamma_{\rm fast} = 5600\,{\rm s}^{-1}$). Our result is also consistent with Eq.~(\ref{eq:Gammaslow}), which predicts $\Gamma_{\rm slow} = 150\,{\rm s}^{-1}$.
 
\subsubsection{Resonance crossover}

\begin{figure}
\centering
\includegraphics[trim=0 6 0 0,clip,width=1\columnwidth]{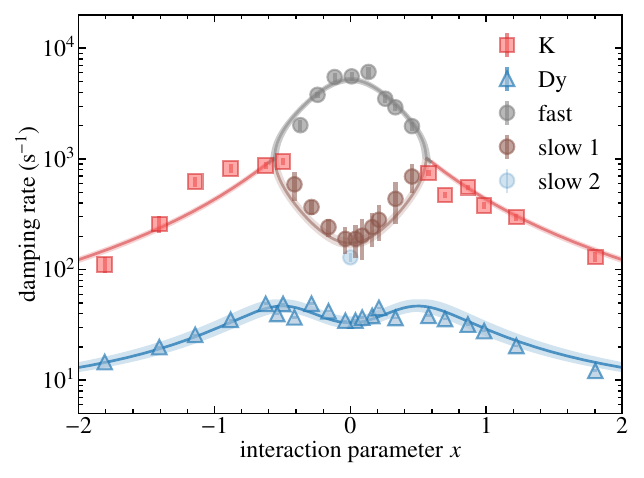}
\caption{Damping rates of all dipole modes across the resonance. The symbols represent the experimental measurements and the solid lines show the result of a combined fit to the data. While, close to resonance, the Dy crossover mode (blue triangles) makes a smooth transition into the hydrodynamic regime, the K mode (red squares) loses its oscillatory character and splits into two exponentially damped modes (circular symbols).
All measurements have been taken with time-of-flight imaging, with the exception of the single data point for the slow mode (labeled `slow 2' ), for which we have applied {\it in situ} imaging.}
\label{fig:FastAndSlowMode}
\end{figure}

We now consider the damping behavior of the dipole modes across the resonance. The experimental data in Fig.~\ref{fig:FastAndSlowMode} 
show how, for strong interactions, the oscillating K mode splits into the two damping modes, while the Dy crossover modes exhibits the continuous behavior already discussed in Sec.~\ref{sec:Dycrossovermode}. To analyze the data we carry out a combined fit to the damping rates of the observed modes in the same way as done for the crossover mode in Sec.~\ref{sssec:fitanalysis}. The fit is carried out with only two free parameters,
$\beta^*$ for the overall dissipative coupling strength and $x_c$ for the effective finite-momentum limitation; see Eq.~(\ref{eq:betax}). Here, the reactive coupling constant is set to $\kappa^* = 0$ as the effect is too weak to be resolved in this data set \footnote{In contrast to the fit analysis of the crossover mode data in Fig.~\ref{fig:DymodeFreqAndDamping}, we cannot extract a meaningful value for the reactive coupling constant $\kappa^*$ from the damping rate data of Fig.~\ref{fig:FastAndSlowMode}. This is a result of a combination of different reasons: The larger number ratio $N_{\rm Dy}/N_{\rm K}$ reduces the damping rate asymmetry of the Dy component, statistical uncertainties are larger, since less data points have been taken in the relevant range, and frequency data are not considered here. A fit with $\kappa^*$ as an additional free parameter would produce a small value with a large uncertainty, which would be consistent with both zero effect and the small value that we would expect from the model in App.~\ref{app:kappa}}.
For the weakly damped Dy crossover mode, as in Sec.~\ref{sec:Dycrossovermode}, we take into account a background damping rate of $6\,{\rm s}^{-1}$, which was measured in the absence of interspecies interactions.

The combined fit reproduces the experimental data well and yields the parameter values $\beta^* = 1090(60)\,{\rm s}^{-1}$ and $x_c = 0.45(3)$, corresponding to $\beta_0 = \beta^*/x_c^2 \approx 5400\,{\rm s}^{-1}$ on resonance.
For the critical value of the friction parameter, above which the damping modes appear, we extract $\beta_{\rm cr} = 2100\,{\rm s}^{-1}$. For the conditions of the present set of experiments, this corresponds to $|x| \le 0.56$ ($|a| \ge 1800\,a_0$). This is {rather} close to the set of measurements on the crossover mode analyzed in Sec.~\ref{sssec:fitanalysis}
{and shows that the particular number ratio ($N_{\rm Dy}/N_{\rm K}$ = 4.5 instead of 2.4) does only weakly affect the extension of the resonant range in which the system exhibits hydrodynamic behavior.}

It is worth to note a detail of the fit. Because of the systematic deviations of unclear origin encountered for the slow damping mode, we have given zero weight to the corresponding data points (marked `slow' and `in situ'). The solid curve for the slow mode is derived from the combined fit to the other modes and shows good agreement with the questionable data points. This gives an {\it a posteriori} justification for our measurements on the slow damping mode.

Our measurements clearly demonstrate the overall behavior predicted by our theoretical model as presented in Sec.~\ref{sec:modespectrum}. In particular, we have studied the resonance crossover, where we have observed the splitting of one of the two oscillating modes (K mode) into two non-oscillating exponentially damped modes.

\section{Microscopic view on friction} 
\label{sec:microfriction}

{
So far, we have described the mode damping in terms of the friction parameter $\beta$ as introduced in Sec.~\ref{ssec:model}.  
This parameter accounts for the total friction force between the two components and thus quantifies its effect on the relative center-of-mass motion. Here, in Sec.~\ref{ssec:localfriction}, we consider the local contributions to friction in the inhomogeneous mixture and introduce a microscopic quantity to describe the friction effect locally. We extract trap-averaged values for this microscopic friction coefficient from the measured damping rates reported in the previous section, and we demonstrate the corresponding resonance behavior. In Sec.~\ref{ssec:universality}, we finally focus on the resonant case of unitarity-limited interactions and interpret our results in terms of universality.
}

\subsection{Local friction} \label{ssec:localfriction}

In the inhomogeneous situation of the trapped mixture, the global friction effect can be understood by considering the sum of the local contributions originating from different regions of the trap. 
While the number densities of the two species, $n_1({\bf r})$ and $n_2({\bf r})$, are local quantities, their relative velocity $\delta \dot{y} = \dot{y}_1-\dot{y}_2$ can be considered as uniform under our basic assumption of two rigid clouds with a relative center-of-mass motion. Accordingly, we start with a general ansatz for the position-dependent volume density of the friction force acting between the two species:
\begin{equation}
    f_1({\bf r}) 
    = - \gamma({\bf r}) \, n_1({\bf r}) \, n_2({\bf r})
    \, \delta\dot{y} \, .
    \label{eq:localfriction}
\end{equation}
{In this way, we define a {\it microscopic friction coefficient} $\gamma({\bf r})$ as a local quantity. The basic idea underlying this ansatz is to separate two-body behavior in a thermal gas, which leads to friction proportional to the local number densities, from many-body effects.
For elastic collisions in a thermal gas without degeneracy effects, it is straightforward to show that $\gamma$ is independent of the position. This can be seen from the Boltzmann equation, where position and momentum dependence separate in this case.
In principle, the friction coefficient $\gamma$ allows for a characterization of the {microscopic} physics of interspecies friction, separated from macroscopic properties of the trapped mixture, such as cloud sizes and shapes. 
}

{
In many-body interaction regimes, $\gamma$ will become position-dependent and we have to take into account the inhomogeneity of the mixture to derive a trap-averaged value $\langle \gamma \rangle$.}
After integrating Eq.~(\ref{eq:localfriction}) over the volume and introducing the reduced total mass $M_r$ as defined in App.~\ref{app:eqmotion}, we obtain 
an expression for the global damping rate of the relative motion:
\begin{equation}
    \beta = 
 - \frac{ \delta\ddot{y}}{\delta \dot{y}}
    = \frac{1}{M_r} \int {\rm d}V \, \gamma({\bf r}) \, n_1({\bf r}) \, n_2({\bf r}) \, .
\label{eq:betavsgamma}
\end{equation}
We now introduce a general definition for the trap average of any position-dependent quantity $\chi({\bf r})$ as 
\begin{equation}
     \langle \chi \rangle \equiv 
\frac{1}{n_\Omega}
\int {\rm d}V \, \chi({\bf r}) \, n_1({\bf r}) \, n_2({\bf r})  \, ,
\label{eq:trapav}
\end{equation}
where $n_\Omega \equiv \int {\rm d}V n_1 n_2$ is the overlap density  (App.~\ref{app:overlap}).
Applying this definition to $\gamma({\bf r})$, Eq.~(\ref{eq:betavsgamma}) can be rewritten as a relation between the trap-averaged microscopic friction coefficient and the global friction parameter,
\begin{equation}
    \langle\gamma\rangle=\frac{M_r}{n_\Omega} \, \beta \, .
    \label{eq:gammavsbeta}
\end{equation}


\begin{figure} 
\centering
\includegraphics[trim=0 6 0 0,clip,width=1\columnwidth]{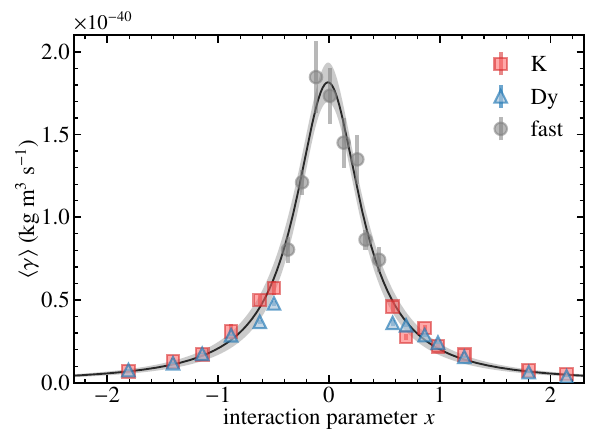}
\caption{Experimental values for the trap-averaged microscopic friction coefficient $\langle \gamma \rangle$ across the resonance, based on the data shown in Fig.~\ref{fig:FastAndSlowMode}. The gray circles, red squares, and blue triangles (shown with standard errors) represent the values extracted from the fast damping mode, Dy crossover mode, and K mode, respectively. The black solid curve is a Lorentzian fit to the fast damping mode only, with the gray-shaded area representing the 1$\sigma$ confidence band.}
\label{fig:microfriction}
\end{figure}

To extract experimental values for $\langle \gamma \rangle$ from our damping rate measurements (Sec.~\ref{sec:dampingmodes}), we first have to convert the damping rates measured for a specific  mode to the corresponding values of $\beta$. For this purpose, we employ our theoretical model, inverting curves such as shown in Fig.~\ref{fig:TheoryCurve}(b).  
{This is particularly straightforward for the fast mode, since its damping rate $\Gamma_{\rm fast}$ approaches $\beta$ in the limit of strong friction (Sec.~\ref{ssec:strongfriction}).}
The fast damping mode also has the great advantage that it acts on a very short timescale of typically much less than a millisecond, {which is about two orders of magnitude faster than the oscillation in the trap}. In this short time, perturbations by losses and heating effects as well as distortions of the cloud can be completely neglected, which minimizes systematic effects compromising our measurements. We therefore expect measurements based on the fast damping mode to be superior to measurements based on the other modes.

Figure \ref{fig:microfriction} shows the values that we obtained for the trap-averaged microscopic friction coefficient $\langle \gamma \rangle$ across the resonance, based on the data points shown in Fig.~\ref{fig:FastAndSlowMode}.
We find that the dependence on the interaction parameter $x$ is well described by a Lorentzian curve. A corresponding fit yields the on-resonance value
\begin{equation}
    \langle \gamma_0 \rangle = 1.8(2) \times 10^{-40} \, {\rm kg\,m^3\,s^{-1}} \, ,
    \label{eq:gamma0}
\end{equation}
which characterizes the strength of the interaction in the unitarity limit. For the width we obtain $x_c = 0.33(3)$, corresponding to $a_c = 3000(300)\,a_0$. This well agrees with the expectation $a_c = 3100\,a_0$ based on Eq.~(\ref{eq:ac}).
We note that our fit only takes into account the fast-mode measurements (eight central data points, gray circles), which we consider to be essentially free of any systematic problems. Without any further fitting, the Lorentzian wings reproduce the damping behavior of the two oscillating modes very well, which nicely confirms the overall consistency of our damping analysis.

\subsection{Universality on resonance} \label{ssec:universality}

On resonance, the $s$-wave scattering length $a$ diverges and no longer represents a relevant length scale for the theoretical description of the system. The resulting universal behavior has been widely discussed in the field of ultracold quantum gases.
For the microscopic friction coefficient, dimensional analysis reveals a natural unit $\hbar/k$, with the inverse wavenumber $1/k$ representing the relevant length scale. The meaning of $k$ depends on the particular regime.

In the {\em thermal regime}, the only relevant length scale is represented by the thermal wavenumber $1/k_{\rm th} = \lambda_{\rm th}/\sqrt{2 \pi} = \hbar/\sqrt{m_r k_B T}$. For the microscopic friction coefficient on resonance we derive the theoretical prediction
\begin{equation}
    \gamma_0 = \frac{8}{3} \sqrt{{2}{\pi}} \frac{\hbar}{k_{\rm th}}
    = \frac{8}{3} \hbar\lambda_{\rm th} \, ,
    \label{eq:gamma0theory}
\end{equation}
which for the experimental parameters of Fig.~\ref{fig:microfriction} yields a unitarity-limited value of $2.4 \times 10^{-40} \, {\rm kg\,m^3\,s^{-1}}$. Note that in the thermal regime, $\gamma$ does not depend on the position and thus $\langle\gamma\rangle = \gamma$. 

Our experimental value of $\gamma_0$, see Eq.~(\ref{eq:gamma0}), lies just 25\% below the theoretical prediction of Eq.~(\ref{eq:gamma0theory}) for the thermal regime.
We believe that this rather small deviation can be attributed to the onset of Pauli blocking in our moderately degenerate sample, 
{which we have verified by extending the numerical simulation approach of Refs.~\cite{Lepers2010nso, Goulko2012bes} to mass-imbalanced mixtures.} 

In the {\em zero-temperature regime} of a degenerate Fermi gas (Fermi energy $E_F$), the inverse Fermi wavenumber $1/k_F$ takes the role of the only relevant length scale.
Consequently, $\hbar/k_F$ represents the natural unit of the microscopic friction coefficient $\gamma$. For the connection between the two limiting regimes, note that $k_{\rm th}/k_F = \sqrt{k_B T/(2 E_F)}$.

For the two-species Fermi-Fermi system, we define 
\cite{Pini2021bmf} the local two-species Fermi wavenumber
\begin{equation}
    {k}_{2F} = [3 \pi^2 (n_1 + n_2)]^{1/3} \, 
\label{eq:k2F}
\end{equation}
and the corresponding local two-species Fermi energy resp.\ temperature
\begin{equation}
    {E}_{2F} = k_B T_{2F} = \frac{\hbar^2 {k}_{2F}^2}{4 m_r} \, .
\end{equation} 
Note that these definitions contain the ones commonly used for unpolarized spin mixtures, representing the special case of balanced masses and populations.

Using units $\hbar / k_{2F}$ for a general description at all temperatures, we write
\begin{equation}
\gamma_0 = \nu(\tau) \frac{\hbar}{k_{2F}} \, .
\label{eq:gam0nu}
\end{equation}
Here we have introduced the dimensionless function $\nu(\tau)$ of the reduced temperature $\tau = T/T_{2F}$ to describe the universal behavior of the microscopic friction effect.
For the thermal regime ($\tau\gg 1$),  Eq.~(\ref{eq:gamma0theory}) can be put in the simple form
\begin{equation}
    \nu = 9.45 \,\tau^{-1/2} \, .
\label{eq:thlimit}
\end{equation}

Let us now interpret our experimental result in relation to the universal function $\nu(\tau)$. Here we have to account for the inhomogeneity of the trapped sample, which makes $k_F$ and indirectly also $\nu$ position dependent. By taking the trap average of Eq.~(\ref{eq:gam0nu}), we obtain
\begin{equation}
     \langle{\gamma}_0\rangle = 
\hbar \, 
\left\langle \nu / k_{2F} \right\rangle \, .
\end{equation}
In a final approximation, we can assume that $\nu$ depends only weakly on the local value of $\tau$, which is motivated by the temperatures being too high for a phase transition to a superfluid  \cite{Pini2021bmf} and by the rather smooth temperature dependence observed in previous experiments in a similar regime \cite{Sommer2011ust, Valtolina2017etf}. 
Thus regarding $\nu$ as a constant factor, we can take it out of the trap-average integral and derive an effective value
\begin{equation}
    \tilde{\nu} = \frac{\langle \gamma_0 \rangle}{\langle 1/k_{2F} \rangle} = 7.8(9) 
    \, ,
    \label{eq:nueff}
\end{equation}
which characterizes the situation in our trap with an effective reduced temperature $\tilde{\tau} = T/\langle T_{2F} \rangle \approx 1.5$, where $\langle T_{2F} \rangle = 85\,$nK is the trap-averaged Fermi temperature.

The present experimental result is indeed close to the thermal limit as described by Eq.~(\ref{eq:thlimit}). It just 
represents one data point to explore or test the universal function $\nu(\tau)$, but in view of future experiments it serves as an example for a general way to characterize friction in a resonant two-component Fermi gas on a microscopic level.

Regarding previous work, the {\em macroscopic} effect of friction between two resonantly interacting fermionic clouds has been investigated by groups at MIT \cite{Sommer2011ust} and at LENS \cite{Valtolina2017etf} employing spin mixtures of $^6$Li. They studied the relative center-of-mass motion and reported measurements of the spin drag rate $\Gamma_{\rm sd}$, which according to their definition corresponds to the friction parameter $\beta$ as defined in our work.
For this quantity of interest, $E_F/\hbar$ represents the natural unit (where $E_F$ is the local Fermi energy in the center of the trap \cite{Sommer2011ust} or the global Fermi energy characterizing the harmonic trap \cite{Valtolina2017etf}). The MIT group reported normalized values of $\hbar \Gamma_{\rm sd}/E_F \approx 0.1$ in a temperature range between $T/T_F = 0.3$ and $3$, while the measurements at LENS provided somewhat higher values varying from $\hbar \Gamma_{\rm sd}/E_F \approx 0.1$ to $0.4$ in a temperature range between $T/T_F = 0.1$ and $0.7$.

In our work, the macroscopic drag effect is described by the parameter $\beta$, which corresponds to the spin drag coefficient. For a reduced temperature $T/{T}_F(0) \approx 1.0$ in the center of the trap, we have measured (data of Fig.~\ref{fig:FastAndSlowMode}) an on-resonance value of $\beta \approx 5400\,{\rm s}^{-1}$ , which corresponds to a normalized value of $\hbar \beta/{E}_{F}(0) \approx 0.34$. This result lies in the range of the measurements in spin mixtures, which have explored essentially the same interaction regime. However, a more accurate and detailed comparison is hampered by the complications related to the inhomogeneity of the harmonically trapped mixtures. Here, a comparison on the level of the microscopic friction coefficient would be highly desirable.

\section{Conclusions and Outlook}\label{sec:conclusion}

In a comprehensive set of experiments, we have investigated the complete dipole-mode spectrum in a two-species fermionic mixture of $^{161}$Dy and $^{40}$K, which features mass imbalance and a widely tunable interspecies interaction. 
While, in the non-interacting case, the two components exhibit individual oscillations with largely different frequencies, dissipative coupling at increasing interaction strength induces damping and finally a crossover into the hydrodynamic regime, leading to a variety of interesting effects. The two oscillating modes exhibit large frequency shifts and, beyond a critical coupling strength in the hydrodynamic regime, one of these modes loses its oscillating character and splits into two purely exponential damping modes.

The most evident signature of hydrodynamics is locking of the individual oscillations to a joint oscillation, in which both species oscillate together with the same frequency. While this effect has been demonstrated in previous work on spin \cite{Gensemer2001tfc}, isotopic \cite{Delehaye2015cva}, or species mixtures \cite{Ferlaino2003doi, Finelli2024ula} with uncoupled frequencies lying rather close to each other (difference at most a few 10\%), our experiments have been carried out in a more extreme regime with individual frequencies differing by a relatively large factor (frequency ratio of $\sim$3).  We have measured the mode frequencies, damping rates along with their relative amplitudes and phases, which fully characterizes the resonance crossover. {These results also reveal an effect beyond 
the mere effect of elastic scattering and resulting friction, demonstrating demonstrating an asymmetry in the damping rate and phase difference, which can be attributed to the reactive mean-field interaction.} 

Exceeding a critical strength of the interspecies friction, one of the two oscillating modes disappears by splitting into two purely exponential damping modes. The resulting slow damping mode essentially corresponds to the overdamped spatial motion in position space, while the fast mode represents the rapidly damped relative motion in momentum space. We have observed both modes and measured their damping rates in the hydrodynamic crossover. We find that the latter mode is largely insensitive against experimental imperfections, as unwanted effects such as losses and heating or dynamics related to the shape of the cloud appear on a much longer time scale.

All our observations can be understood in terms of a rather simple model from the literature \cite{Gensemer2001tfc, Ferlaino2003doi, Chiacchiera2010doi, Asano2020doo}, which considers the one-dimensional relative motion of the clouds' mass centers. The interaction coefficients describing the dissipative and reactive coupling can be determined experimentally by fitting the model predictions to the measurements. We found the extracted values to be consistent with theoretical calculations based on collisional dynamics and the mean-field interaction. In this approach, the clouds are considered as rigid mass distributions, which is a benefit of the excellent spatial overlap between the two species in our particular two-species system. 

Our interpretation of the {observed interspecies friction goes beyond a description of the macroscopic drag effect} between the two components, which is complicated by the inhomogeneous trap environment. We have introduced a microscopic friction coefficient as a local quantity and extracted corresponding {trap-averaged} values from our measurements of mode damping in the strongly interacting regime. We have also shown how the result obtained on resonance, where the scattering length diverges, can be interpreted in terms of universal behavior in the thermal or degenerate Fermi regime. This facilitates a comparison of experiments carried out with a wide range of different two-species systems in the resonantly interacting regime.
 
In view of future experiments on strongly interacting two-species quantum mixtures, measurements of dipole modes offer a versatile tool-box to investigate phenomena related to hydrodynamic behavior. In particular, the fast damping mode, demonstrated in this work, constitutes a promising new tool for precision experiments related to the fast dissipative dynamics of interspecies friction. 
{It represents an experimental implementation of the general idea \cite{Carlini2021sda} to study the response of a mixture to a sudden, species-selective perturbation on a short timescale.
This method} may even be applied to quantum gases with resonantly interacting bosonic components \cite{Varenna2022book}, where the much shorter lifetimes will not permit the observation of oscillation modes. 

{In ongoing experiments on the Dy-K system, after various improvements of the preparation process, we are now able to produce colder mixtures near the Feshbach resonance~\footnote{As the main problem we identified heating when the deeply degenerate mixture is transferred from the low magnetic field (250\,mG), where evaporative cooling is performed, to higher fields near the 7.3\,G Feshbach resonance, which is used for the present experiments. This heating is due to a combination of detrimental effects from interspecies~\cite{Ye2022ool} and intraspecies \cite{Soave2022lff} Feshbach resonances.}. 
Advancing in near-future work into more deeply degenerate regimes,
we can expect substantial manifestations of additional effects beyond collisional hydrodynamics to become observable.
With deeper cooling, increased Pauli blocking effects will first reduce the damping rates, before pair correlations and eventually the formation of a superfluid will strongly affect the mode frequencies \cite{Tajima2020sdm}.}

In a broader sense, our work on two-species dipole modes generalizes previous work on fermionic spin mixtures~\cite{Sommer2011ust, Sommer2011sti}, which addressed fundamental transport properties in a strongly interacting gas along with their quantum limits. In this context, the microscopic friction coefficient introduced in our work will represent an important building block for understanding the full dynamics in future investigations.

\section*{Acknowledgments}

We thank M.~Zaccanti, S.~Finelli and the members of the Li-Cr team at LENS for many stimulating discussions on fermionic quantum mixtures with resonant interactions. 
We acknowledge inspiring discussions with O.~Goulko and M.~Zwierlein.
We also wish to thank L.~Absil for discussions on the manuscript and all members of the ultracold groups in Innsbruck for a stimulating environment and many insights related to the broad field of quantum-gas mixtures.
The project has received funding from the European Research Council (ERC) under the European Union’s Horizon 2020 research and innovation programme (Grant Agreement No. 101020438 - SuperCoolMix). We further acknowledge support by the Austrian Science Fund (FWF) within the Doktoratskolleg
ALM (W1259-N27).
ZXY acknowledges support by the National Natural Science Foundation of China (No.~U23A6004) and the Fundamental Research Program of Shanxi Province (No.~202403021221001).

\appendix

\section{Equations of motion and solution}
\label{app:eqmotion}

The two components are treated as rigid masses $M_i = N_i m_i$, where $N_i$ is the atom number of each species and $m_i$ represents the respective atomic mass. The total force on each of the two clouds is a sum of three contributions: (1) The external, species-dependent force exerted by the trap is characterized by the corresponding trap frequencies $\omega_1$ and $\omega_2$, (2) the dissipative friction force is proportional to the relative velocity of both components with the coefficient $B$, and (3) the reactive force is proportional to the relative displacement, which is assumed to be small compared with the cloud sizes. This force represents the attraction or repulsion between the two components and is quantified by a spring constant $K$. 

The equations of motion for the mass centers of the two components can be written as 
\begin{subequations}
  \begin{eqnarray}
     M_1 \ddot{y}_1 & = & - M_1 \omega_1^2 y_1 - B (\dot{y}_1 - \dot{y}_2) - K (y_1 - y_2)  \makebox[6mm][c]{} \\
     M_2 \ddot{y}_2 & = & - M_2 \omega_2^2 y_2 - B (\dot{y}_2 - \dot{y}_1) - K (y_2 - y_1) 
\end{eqnarray}
\label{eq:macroscopic}
\end{subequations}     
In the explicit form, the system of two coupled, second-order linear differential equations reads:
\begin{subequations}
  \begin{eqnarray}
     \ddot{y}_1 & = & -\omega_1^2 y_1 - \beta_1 (\dot{y}_1 - \dot{y}_2) - \kappa_1 (y_1 - y_2) \\
    \ddot{y}_2 & = & - \omega_2^2 y_2 - \beta_2 (\dot{y}_2 - \dot{y}_1) - \kappa_2 (y_2 - y_1)
\end{eqnarray}
\end{subequations}     
where $\beta_i = B / M_i$ and $\kappa_i = K / M_i$.

Expressing the system of equations in matrix form
\begin{equation}
\left( \begin{array}{cccc}
\dot{y}_1 \\
\dot{y}_2 \\
\ddot{y}_1 \\
\ddot{y}_2 \\
\end{array}\right)
=
\left( \begin{array}{cccc}
0 & 0 & 1 & 0 \\
0 & 0 & 0 & 1 \\
-\omega_1^2 - \kappa_1 & \kappa_1 & -\beta_1 &\beta_1 \\
\kappa_2          & -\omega_2^2-\kappa_2 & \beta_2 & -\beta_2 \\
\end{array}\right)
\left( \begin{array}{cccc}
{y}_1 \\
{y}_2 \\
\dot{y}_1 \\
\dot{y}_2 \\
\end{array}\right)
\end{equation}
facilitates a solution by standard numerical routines, which finally yields the complex-valued eigenfrequencies and eigenvectors of the different modes.

For the discussion in the main text, we introduce the reduced total mass $M_r = M_1 M_2 / (M_1 +M_2)$ 
and we define the mass factors $q_1 = M_r/M_1$ and $q_2 = M_r/M_2$. 
The two parameters
\begin{equation}
\beta \equiv \frac{B}{M_r} = \frac{\beta_1}{q_1} = \frac{\beta_2}{q_2}
\end{equation}
and
\begin{equation}
\kappa \equiv \frac{K}{M_r} = \frac{\kappa_1}{q_1} = \frac{\kappa_2}{q_2} \, 
\end{equation}
then characterize the strength of the dissipative and the reactive interaction between the two clouds.

\section{Spatial overlap}
\label{app:overlap}

\begin{figure}
\centering
\includegraphics[trim=0 15 0 0,clip,width=1\columnwidth]{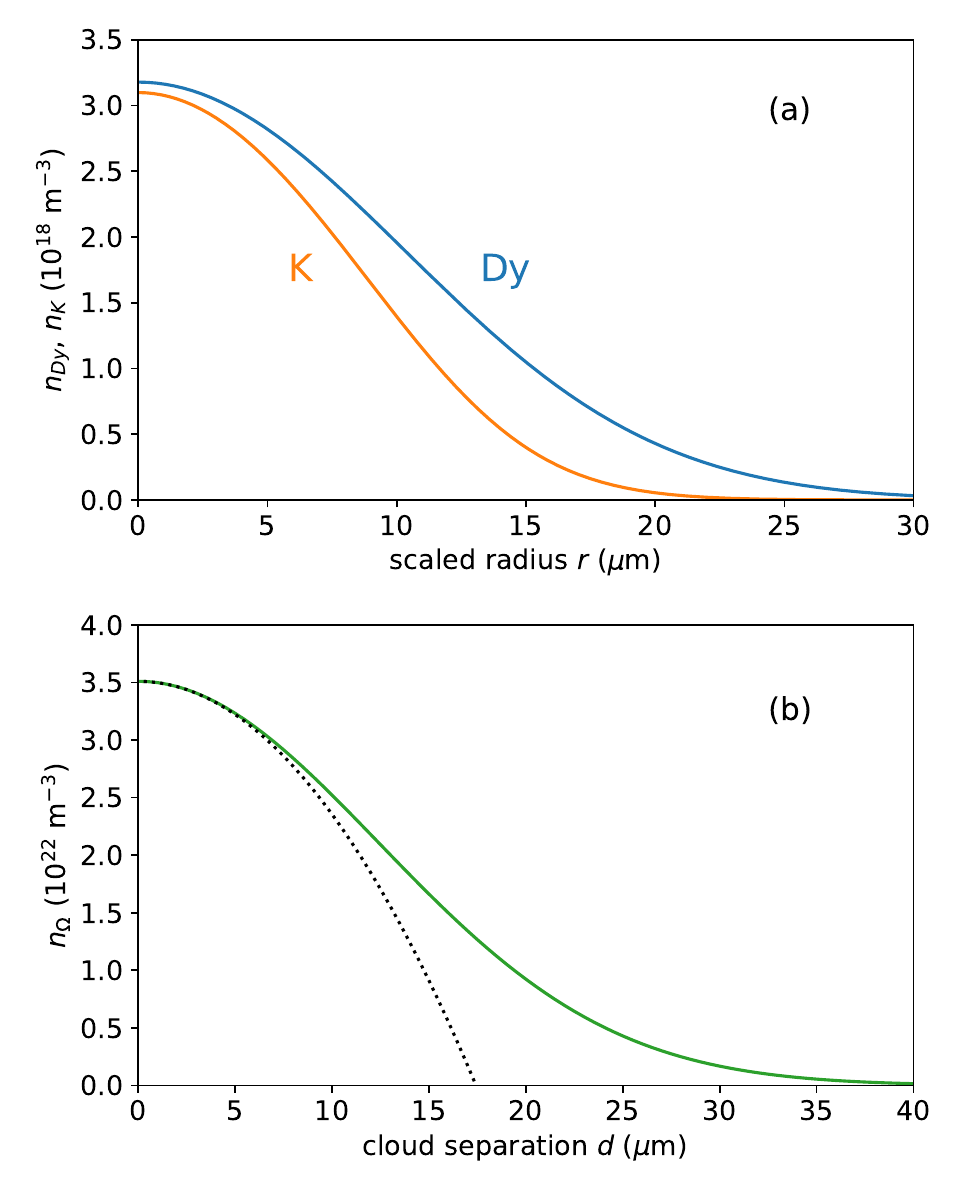}
\caption{Species overlap in the mixture. (a) Spatial profiles for typical experimental conditions $N_{\rm Dy} = 5\times10^4$, $N_{\rm K} = 2\times10^4$, $T = \qty{100}{nK}$ in a spherically symmetric trap with mean frequencies $\bar{\omega}_{\rm Dy}/2\pi = \qty{37}{Hz}$ and $\bar{\omega}_{\rm K}/2\pi = \qty{112}{Hz}$. The reduced temperatures $T_{\rm Dy}/T_F^{\rm Dy} = 0.84$ and $T_{\rm K}/T_F^{\rm K} = 0.38$ correspond to moderate Fermi degeneracy.
(b)~Overlap integral $n_\Omega$ as a function of a spatial displacement between the cloud centers (solid line) with a second-order Taylor expansion (dotted line).
}
\label{fig:overlap}
\end{figure}

Throughout this work, we model the Dy and K clouds as rigid mass distributions with spatial equilibrium shapes corresponding to non-interacting Fermi gases. Figure~\ref{fig:overlap}(a) shows the number density profiles $n_{\rm Dy}({r})$ and $n_{\rm K}({r})$ calculated for both species under typical experimental conditions in our harmonic trap. The profiles exhibit near spatial matching and thus a good overlap between both components. The Dy cloud is somewhat larger than the K cloud, but only a small fraction in the outer region does not show any significant overlap with K atoms.

To quantify the spatial overlap between the two components, we define the {\em overlap density}
\begin{equation}
    n_{\Omega} = \int {\rm d}V n_{\rm Dy}({\bf r}) \, n_{\rm K}({\bf r}) \, .
\label{eq:overlapgen}
\end{equation}
For spherical symmetry the volume integral simplifies to
\begin{equation}
    n_{\Omega} 
    = 4 \pi \int_0^\infty dr \, r^2 \, n_{\rm Dy}({r}) \, n_{\rm K}({r}) \, .
\label{eq:overlapsymm}
\end{equation}
For the interpretation of the overlap density, note that $n_\Omega / N_{K} = \langle n_{\rm Dy} \rangle_{\rm K}$ represents the mean density of Dy atoms as `seen' by the K atoms (and vice versa).

A spatial displacement $d$ between the two clouds reduces the overlap density. In Fig.~\ref{fig:overlap}(b), we show $n_{\Omega}(d)$ (solid line) as calculated numerically from Eq.~(\ref{eq:overlapgen}).
For small displacements, a Taylor expansion can be applied:
\begin{equation}
n_{\Omega}(d) = n_{\Omega}(0) + \frac{1}{2} d^2  n''_{\Omega}(0) \, .
\label{eq:overlap2}
\end{equation}
For the conditions of Fig.~\ref{fig:overlap}, we obtain $n_{\Omega} = 3.51\times 10^{22}\,{\rm m}^{-3}$ and $n''_{\Omega}(0) = -2.31 \times 10^{32}\,{\rm m}^{-5}$ by numerically evaluating the overlap integrals.
The dotted curve in Fig.~\ref{fig:overlap}(b) illustrates that the second-order expansion remains a good approximation up to displacements of about \qty{10}{\mu m}, which corresponds to the linear regime of the reactive force between the two clouds.

We finally have to take into account the anisotropy of our cigar-shaped harmonic trap, which has radial frequencies $\omega_x = \omega_y$ ($2\pi \times 59$\,Hz for Dy, $2 \pi \times 177$\,Hz for K), but about four times lower axial frequencies $\omega_z$. We introduce the scaled radius
\begin{equation}
    r = \sqrt{(\alpha x)^2 + (\alpha y)^2 + (\alpha^{-2} z^2)} 
\end{equation}
with a scaling parameter $\alpha \equiv (\omega_x/\omega_z)^{1/3} \approx 1.6$. Using this definition, the local density approximation allows us to map the situation onto the spherically symmetric case and to apply the above approach. Finally the results can be scaled back to the anisotropic trap. This has no effect on $n_{\Omega}(0)$, but $n''_{\Omega}(0)$ has to be multiplied by the factor $\alpha^2 \approx 2.5$. Moreover, the linear regime shrinks by the factor $\alpha$ to displacements less than about \qty{6}{\mu m}.
 

\section{Modelling the interspecies interaction}
\label{app:cloudinteraction}

Here we describe the interaction between the two species in a collisional mean-field model. We introduce the functions $\beta(x)$ and $\kappa(x)$ used in the main text to describe the resonance behavior of the dissipative and reactive interaction between the two species. Within the approximation of rigid mass distributions, we neglect any interaction-induced changes of the cloud profiles.

\subsection{Dissipative interaction (friction)}
\label{app:beta}

We first consider elastic collisions with a fixed $s$-wave scattering cross section $\sigma = 4 \pi a^2$. For the moderate Fermi degeneracy in our experiments, we neglect Pauli blocking effects on the collisions and we assume a mean relative velocity of $\bar{v}_{\rm rel} = \sqrt{8 k_B T/(\pi m_r})$, which corresponds to the thermal limit, where $m_r$ is the individual particle reduced mass. With this assumption, we obtain 
\begin{equation}
    \Gamma_{\rm coll} = n_{\Omega} \sigma \bar{v}_{\rm rel} \, 
    \label{eq:collrate}
\end{equation}
for the total elastic collision rate \cite{Mosk2001mou};
here $n_{\Omega}$ respresents the overlap density as discussed in Appendix~\ref{app:overlap}.
Integrating over all collisions in the mixture we obtain the relation to the damping rate for the relative motion of the two species, for which we have introduced the friction parameter
\begin{equation}
    \beta = \frac{4}{3} \frac{m_r}{M_r} \Gamma_{\rm coll} \, .
    \label{eq:betacollrate}
\end{equation}
We note that this result is in agreement with the model presented in Ref.~\cite{Gensemer2001tfc}. We have also verified the relation between $\beta$ and $\Gamma_{\rm coll}$ independently within Monte-Carlo simulations of the damping of the relative interspecies motion. 

Combining Eqs.~(\ref{eq:collrate}) and (\ref{eq:betacollrate}), we obtain the general relation between the friction coefficient and the scattering cross section,
\begin{equation}
    \beta = \frac{4}{3} \frac{m_r}{M_r} n_{\Omega}  \bar{v}_{\rm rel} \times \sigma \, .
    \label{eq:betasigma}
\end{equation}
In the resonance regime, where $\sigma$ depends on the relative momentum, this equation remains valid, provided that
a properly defined effective cross section $\sigma_{\rm eff}$ is introduced. We note that, staying with this definition, the collision rate $\Gamma_{\rm coll}$ according to Eq.~(\ref{eq:collrate}) acquires an additional factor of $2$, and the prefactor in Eq.~(\ref{eq:betacollrate}) has to be changed to $2/3$. One should also bear in mind that Eqs.~(\ref{eq:collrate}-\ref{eq:betasigma}) have been derived under the assumption of thermal velocity distributions. Deviations remain rather small under our present experimental conditions, but Fermi degeneracy will play a much more important role in future experiments at lower temperatures.

We model the resonance behavior of the cross section based on the imaginary part of the $s$-wave scattering amplitude, considering the unitarity-limitation imposed by the finite collision momenta.
We neglect effective-range corrections to the scattering amplitude and thus momentum-dependent resonance shifts, which is a good approximation for the sufficiently broad Feshbach resonance employed in this work. By integration over all collisional momenta, we obtain an effective cross section, which can be well approximated by a Lorentzian
\begin{equation}
    \sigma_{\rm eff} = \frac{4 \pi a^2}{1+a^2/a_c^2} \, ,
    \label{eq:sigmareso}
\end{equation}
where the length $a_c$ accounts for the finite-momentum limitation. From the theory of resonant scattering in a thermal gas, we can derive the approximation 
\begin{equation}
    a_c \approx 1.21\,\frac{\lambda_{\rm th}}{2 \pi} \, ,
    \label{eq:ac}
\end{equation}
where $\lambda_{\rm th} = \sqrt{2 \pi \hbar^2/(m_r k_B T)}$ represents the thermal de~Broglie wavelength associated with the relative motion. While Eq.~(\ref{eq:sigmareso}) reproduces the zero-temperature limit $4\pi a^2$ exactly, we note that within the Lorentzian approximation the on-resonance cross section $4\pi a_c^2 \approx 0.47\,\lambda_{\rm th}^2$ turns out to be about 7\% lower than the exact result $\lambda_{\rm th}^2/2$. This minor deviation is not relevant for the present experiments.

Combining the above Eqs.~(\ref{eq:collrate}-\ref{eq:sigmareso}) and using the dimensionless parameter $x = 1000\,a_0/a$, defined in Eq.~(\ref{eq:definex}) in the main text to characterize the interaction strength across the resonance, we obtain the function
\begin{equation}
    \beta(x) = \frac{\beta^*}{x^2 + x_c^2} \, ,
\end{equation}
which describes the resonance behavior of the dissipative interaction (friction) between the two species.
For the overall strength of this interaction, the collisional model predicts 
\begin{equation}\label{eq:betastar}
    \beta^* = \frac{16 \pi } {3} \frac{ m_r } { M_r} n_{\Omega} \bar{v}_{\rm rel}  \,  \times (1000\,a_0)^2 \, .
\end{equation}
For the conditions of our experiments on the Dy crossover mode reported in Sec.~\ref{sec:Dycrossovermode} the relevant parameter values with estimated uncertainties are
$m_r/M_r = 3.7(4) \times 10^{-5}$, $n_{\Omega} = 4.8(5) \times 10^{22}\,\textrm{m}^{-3}$, and $\bar{v}_{\rm rel} = 8.1(8)\,\textrm{mm/s}$. 
This results in a parameter value of $\beta^* =  680(120)\,\mathrm{s}^{-1}$, where the
rather large uncertainty of nearly 20\% reflects our limited knowledge of the experimental conditions.

\subsection{Reactive interaction (mean field)}
\label{app:kappa}

The total potential energy resulting from the mean-field interaction between both species is given by the general expression
\begin{equation}
    U_{\rm mf} = g\,n_{\Omega} \, ,
\end{equation}
where $g = (2 \pi \hbar^2 a)/m_r$ represents the coupling constant.
With the overlap density according to Eq.~(\ref{eq:overlap2}), which
describes the effect of small displacements between the two clouds, we obtain
\begin{equation}
    U_{\rm mf}(d) = U_{\rm mf}(0) + \frac{1}{2} g \, \alpha^2 n''_{\Omega}(0) \, d^2 \, .
\end{equation}
The second term ($\propto d^2$) quantifies the reactive effect in a harmonic approximation, with $g\, \alpha^2 n''_{\Omega}(0)$ representing the spring constant. The latter corresponds to the coefficient $K = M_r \kappa$ in Eq.~(\ref{eq:macroscopic}). We thus obtain
\begin{equation}
    \kappa = \frac{g\, \alpha^2}{M_r}  n''_{\Omega}(0) \,
    \label{eq:kappaoverlap}
\end{equation}
which is positive for attractive interaction.

To model the unitarity-limited resonance behavior in a way analogous to Eq.~(\ref{eq:sigmareso}), we introduce the effective coupling constant
\begin{equation}
    g_{\rm eff} = \frac{2 \pi \hbar^2}{m_r} \frac{a}{1+a^2/a_c^2} \, ,
    \label{eq:geff}
\end{equation}
which results from real part of the $s$-wave scattering amplitude in the resonance regime.

Combining the above Eqs.~(\ref{eq:kappaoverlap} and \ref{eq:geff}) and again characterizing the interaction by the parameter $x = 1000\,a_0/a$, we obtain the function
\begin{equation}
    \kappa(x) = -\frac{\kappa^* x}{x^2 + x_c^2} \, ,
\end{equation} 
where
\begin{equation}\label{eq:kappastar}
    \kappa^* = - \frac{2\pi\hbar^2 \alpha^2 n''_{\Omega} }{M_r m_r} \,  \times 1000\,a_0 \, .
\end{equation}
For the experimental conditions of the experiments described in Sec.~\ref{sec:Dycrossovermode}, 
$M_r = 8.7(9) \times 10^4\,\textrm{u}$,
$\alpha^2 = 2.6(3)$, and 
$n''_{\Omega}(0) = -3.1(3) \times 10^{32}\,{\rm m}^{-5}$, 
we finally obtain the parameter value
$\kappa^* = 3.8(7)\times10^4\,\textrm{s}^{-2}$.
Regarding the uncertainty we note that a substantial additional  uncertainty may arise from an interaction-induced deformation of the clouds, which we have neglected in our simple model and which presumably can have a large effect on the derivative $n''_{\Omega}$.


%

\end{document}